\documentclass[preprint,12pt,authoryear]{elsarticle}

\usepackage{amssymb}
\usepackage{amsmath}

\newcommand{\ele}[1]{\textcolor{orange}{#1}}
\newcommand\elesout{\bgroup\markoverwith{\ele{\rule[0.5ex]{2pt}{1.0pt}}}\ULon}

\newcommand{\yv}{\mathbf{y}}

\newcommand{\bv}{\mathbf{b}}

\newcommand{\fv}{\boldsymbol{f}}

\newcommand{\betav}{\boldsymbol{\beta}}
\newcommand{\varepsilonv}{\boldsymbol{\varepsilon}}
\newcommand{\zerov}{\mathbf{0}}
\newcommand{\ev}{\mathbf{e}}

\newcommand\blfootnote[1]{%
  \begingroup
  \renewcommand\thefootnote{}\footnote{#1}%
  \addtocounter{footnote}{-1}%
  \endgroup
}

\DeclareMathOperator*{\argmin}{arg\min}

\DeclareMathOperator{\cardO}{|\mathcal{O}|}
\DeclareMathOperator{\cardOk}{|\mathcal{O}_k|}
\DeclareMathOperator{\fasynt}{\breve{\mathbf{f}}_{\cardO}}
\DeclareMathOperator{\betaasynt}{\breve{\mathbf{\betav}}_{\cardO}}

\newtheorem{proposition}{Proposition}[section]

\usepackage{glossaries}
\usepackage{verbatim}
\usepackage{xcolor}
\usepackage{hyperref} 
\usepackage{array}
\usepackage{mathrsfs} 

\usepackage{xr}
\makeatletter

\newcommand*{\addFileDependency}[1]{
\typeout{(#1)}
\@addtofilelist{#1}
\IfFileExists{#1}{}{\typeout{No file #1.}}
}\makeatother
\newacronym{pde}{PDE}{Partial Differential Equation}
\newacronym{fpirls}{FPIRLS}{Functional Penalized Iterative Reweighted Least Squares}
\newacronym{arpa}{ARPA}{Agenzia Regionale per la Protezione dell'Ambiente}
\newacronym{mestpde}{\texttt{MEST-PDE}}{Mixed-Effect Spatio-Temporal Regression with Partial Differential Equation regularization}
\newacronym{rmse}{RMSE}{Root Mean Squared Error}

\journal{Nuclear Physics B}

\begin{document}

\begin{frontmatter}

\title{Modeling group heterogeneity in spatio-temporal data via physics-informed regression \blfootnote{This is an accepted manuscript of an article published by Elsevier in Spatial Statistics. The final published version is available online at: https://doi.org/10.1016/j.spasta.2026.101022}}

\author[1]{Marco F. De Sanctis}
\author[2]{Eleonora Arnone} 
\author[1]{\\Francesca Ieva} 
\author[1]{Laura M. Sangalli\corref{cor1}}

\cortext[cor1]{Corresponding author: laura.sangalli@polimi.it}
\affiliation[1]{organization={MOX, Dipartimento di Matematica, Politecnico di Milano},
            addressline={\\Piazza Leonardo Da Vinci 32}, 
            city={Milano},
            postcode={20133}, 
            country={Italy}}
\affiliation[2]{organization={Dipartimento di Management, Università degli Studi di Torino},
            addressline={\\Corso Unione Sovietica, 218 bis}, 
            city={Torino},
            postcode={10134}, 
            country={Italy}}            

\begin{abstract}
We propose a physics-informed semiparametric framework for modeling spatio-temporal data with group structure. The approach extends classical mixed-effects regression by incorporating a nonparametric space–time component, regularized through a partial differential equation to encode the underlying physical dynamics, while random effects capture group-specific variability.  Estimation is carried out via a two-step procedure based on a functional extension of the Iteratively Reweighted Least Squares algorithm. We establish asymptotic properties of both fixed and random effect estimators and we assess the performance of the method through simulation studies against state-of-the-art alternatives.  The proposed framework is applied to hourly nitrogen dioxide data over Lombardy (Italy), where random effects account for measurement heterogeneity across monitoring stations with different sensor technologies, demonstrating its effectiveness in capturing both physical dynamics and group heterogeneity.
\end{abstract}


\end{frontmatter}


\section{Introduction}
\label{Section:intro}
\begin{figure}[bt]
\hspace{-0.5cm}
\begin{tabular}{p{0.0\textwidth}p{.3\textwidth}p{.3\textwidth}p{.3\textwidth}}
& \qquad \quad \quad \small{08:00} &  \quad \qquad \quad \small{16:00} &  \quad \qquad \quad \small{21:00} \\
&\includegraphics[width=.3\textwidth]{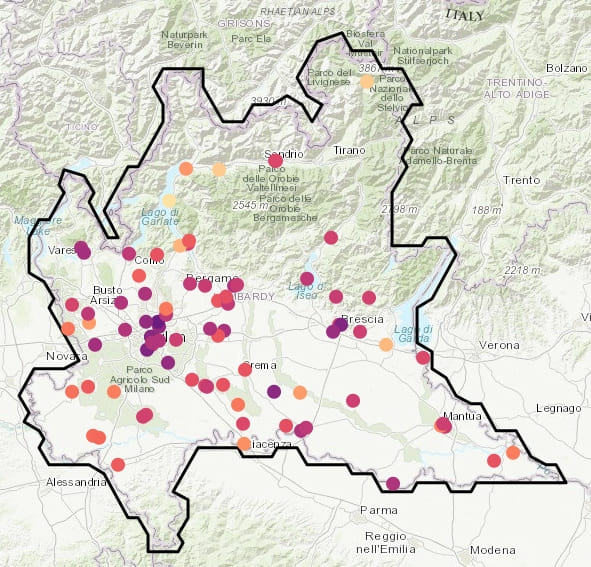}
&\includegraphics[width=.3\textwidth]{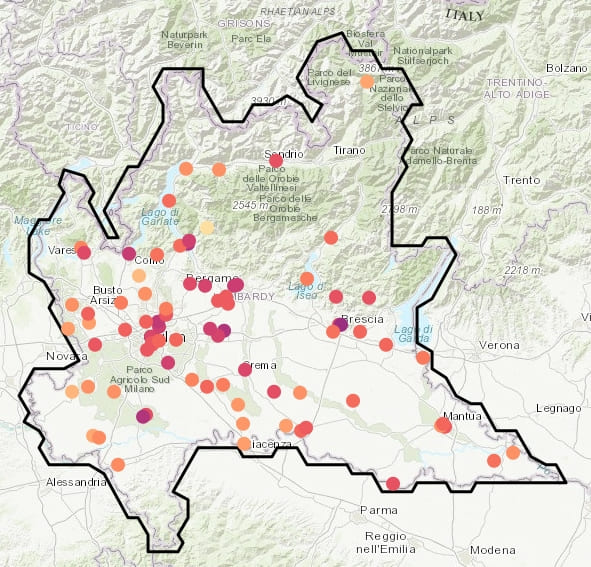}
&\includegraphics[width=.3\textwidth]{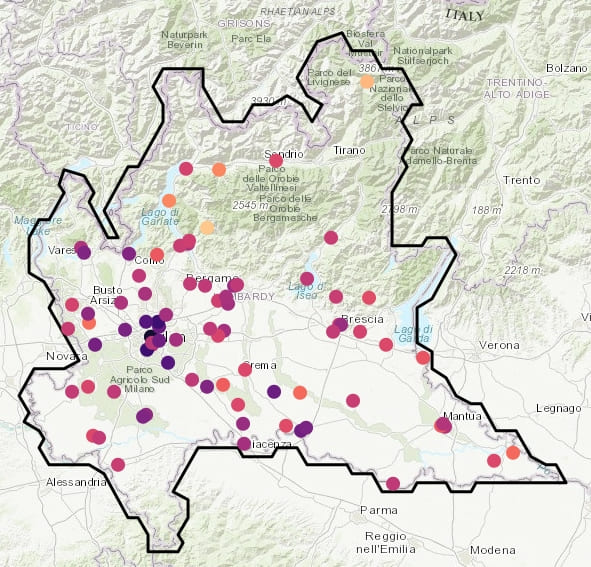}
\end{tabular}%

\centering
\includegraphics[width=0.5\textwidth]{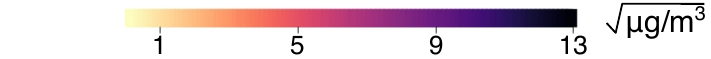} \\ 

\vspace*{0.1cm}
\small{\hspace{0.1\textwidth} Hourly NO$_2$ time profiles \hspace{0.22\textwidth} Sensors types}\\
\includegraphics[width=0.98\textwidth]{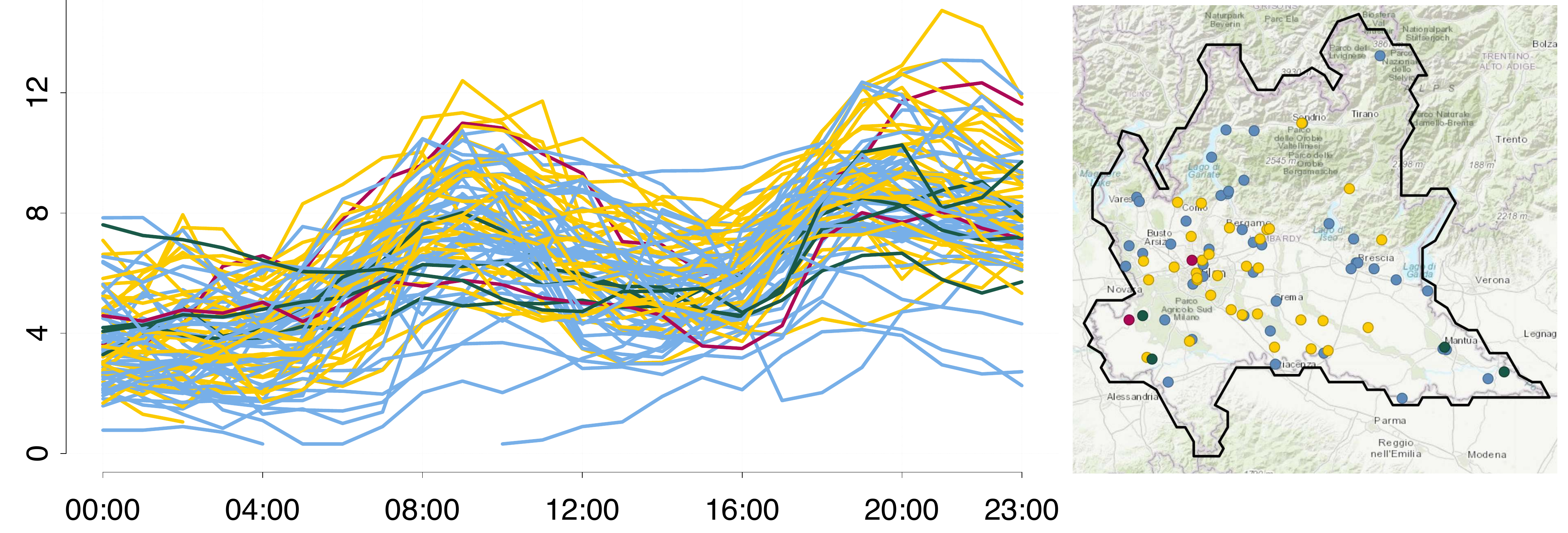}\\
\includegraphics[width=0.5\textwidth]{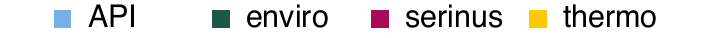} \\

\caption{Top panel: spatial distribution of square-root NO$_2$ concentrations in Lombardy at three representative hours of the day (08:00, 16:00, and 21:00). Bottom panel: hourly temporal profiles of square-root NO$_2$ concentrations across ARPA monitoring stations on $15$ January $2019$ (left); spatial distribution of sensor technology types across the region (right).}
\label{fig:data}
\end{figure}

\begin{figure}[htbp]
\centering
\small{\hspace{-0.075\textwidth}Wind field}\\
\includegraphics[width=0.45\textwidth]{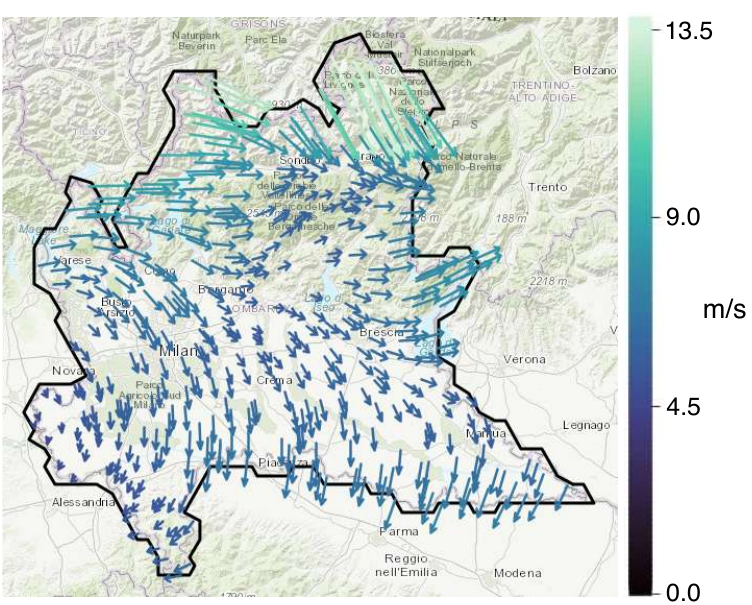}
\caption{Daily average wind vector field over Lombardy on $15$ January $2019$. Wind intensity and direction are derived from data collected at $119$ ARPA monitoring stations.}
\label{fig:wind}
\end{figure}

In this work, we address the modeling of spatio-temporal data exhibiting a group structure, which may arise, for instance, from the use of different measurement technologies. Our focus is on phenomena characterized by complex spatio-temporal patterns driven by external forces, a common feature in many real-world applications, particularly in environmental sciences. An illustrative example is shown in Figure \ref{fig:data}, which reports hourly measurements of nitrogen dioxide (NO$_2$) collected on $15$ January $2019$ by the \gls{arpa} monitoring network in the Lombardy region (Italy). These observations exhibit pronounced temporal variability and sharp spatial gradients, strongly influenced by air circulation. Modeling such processes entails several challenges. First, monitoring sensors differ in technology, design, and calibration systems (see the bottom-right panel of Figure \ref{fig:data}). These differences induce a natural grouping structure, requiring statistical models capable of disentangling group-specific variability from the underlying signal of interest. A second challenge arises from the influence of physical mechanisms, such as wind dynamics and diffusion processes, on pollutant concentrations. Figure \ref{fig:wind} illustrates the wind field observed on the same day as the NO$_2$ measurements. Jointly accounting for group heterogeneity and physically driven dynamics is essential for an accurate characterization of the phenomenon.

In classical linear regression settings, where no spatial or temporal structure is present, group-specific effects are typically modeled through random components, leading to linear mixed-effects models \citep[see, e.g.,][]{pinheiro_SandSplus, galecki2012linear}. A substantial body of work has extended this framework to spatio-temporal regression. In this context, however, random effects are most often introduced to capture residual spatial dependence, rather than to explicitly represent grouping structures within the data. Applications of such spatial random effect models include ordinal data \citep{mullen2008mixed}, compositional data \citep{dibrisco2021}, and environmental data \citep{smith2003spatiotemporal}, with several contributions also addressing spatial confounding \citep[e.g.,][]{khan2022restricted}. More general formulations further extend generalized linear mixed models to additive or nonparametric settings \citep{lin_zhang1999, karcher2001generalized}.
In contrast, a smaller body of literature incorporates random effects in spatio-temporal models specifically to represent group structures, which is also the focus of this work. For instance, \citet{wood2006low} introduced random effects to capture grouping factors while modeling spatial and temporal dependence through low-rank smooths. \citet{yanosky2014spatio} applied this framework to particulate matter data, including site-specific random effects to account for unobserved heterogeneity across monitoring stations, while modeling large-scale trends via smooth functions of geographic and meteorological covariates. Similarly, \citet{sahu2006spatio} modeled fine particulate matter concentrations using a combination of fixed effects and spatio-temporal random components distinguishing between rural and urban areas. More recently, \citet{damatta2025bayesian} proposed a Bayesian spatio-temporal functional model in which random effects represent regional climate regimes.
These contributions highlight the role of random components in flexibly capturing additional sources of variability associated with grouping structures.

A further challenge arises from the influence of complex physical mechanisms, such as wind currents and diffusion processes, on pollutant concentrations. In recent years, the integration of physical information into statistical models has received increasing attention. Among these approaches, a prominent line of research incorporates such information through Partial Differential Equations (PDEs).
Early contributions in this direction include the PDE-based regularization framework for spatial regression proposed by \citet{azzimonti_2014,azzimonti_blood_2015} and further developed by \citet{tomasetto_2024}, with extensions to the spatio-temporal setting introduced by \citet{arnone2019modeling}. More recently, stochastic PDE approaches have been extended to incorporate physically interpretable features such as anisotropy and transport term \citep{CLAROTTO2024, carrizo2022general},  building on the seminal work of \citet{Lindgren_2011} and the literature  reviewed in \citet{lindgren2022spde}. Additional developments include the modeling of non-stationary Gaussian random fields on compact Riemannian manifolds \citep{pereira2022geostatistics} and the construction of physics-informed covariance functions, such as those based on the exponential Boltzmann–Gibbs representation \citep{allard2021linking}.
Alternative approaches integrating PDEs in time-varying settings have also been proposed, including \citet{wikle-2010-TEST}, \citet{richardson-2017}, and \citet{hefley-2017}.

Despite these advances, existing approaches do not explicitly integrate physics-informed modeling with group-structured data.
Building on this perspective, we propose a physics-informed semiparametric mixed-effects regression model that integrates data-driven modeling with the physical dynamics of the underlying process. The model includes fixed and random parametric components to account for covariate effects and group heterogeneity, together with a nonparametric component describing the nonlinear spatio-temporal evolution of the phenomenon.
The estimation of this component is guided by a physics-informed regularization term based on a \gls{pde} that, in the application to NO$_2$ data, encodes pollutant dispersion through an advection term driven by the wind field (see Figure \ref{fig:wind}). This mixed-effects formulation provides two main advantages. First, it enables the modeling of group-specific variability, distinguishing among sensor technologies within the monitoring network. Second, it allows for the separation of measurement noise, arising from instrumentation differences, from the underlying spatio-temporal signal of interest.
From a methodological standpoint, the inclusion of random effects in the objective functional leads to a non-quadratic optimization problem, as the covariance matrix of the random components enters nonlinearly. As a result, iterative procedures are required to solve the estimation problem.

An additional strength of the proposed framework lies in its ability to handle missing data, a common feature of spatio-temporal environmental studies based on sensor networks. Measurements from monitoring stations are often incomplete due to temporary malfunctions, equipment failures, or maintenance interruptions. To address this issue, we formulate the model within the statistical framework of \citet{Arnone2023}, which ensures stable estimation in the presence of missing data.
Moreover, the proposed approach can be applied to data defined over spatial domains with complex geometries. For the spatial discretization of the nonparametric component, we employ a finite element basis, which is well suited to represent phenomena evolving over irregular or non-convex regions, including domains with natural barriers or curved surfaces. This flexibility is particularly important when the physical dynamics of the process are influenced by the geometry of the domain, as in applications involving water bodies with irregular coastlines or biological signals observed on complex three-dimensional structures \citep[see, e.g.,][]{sangalli_spatial_2021, tomasetto_2024, castiglione_2025}.

We investigate the theoretical and empirical properties of the proposed framework. In particular, we derive asymptotic results for both fixed and random effect estimators. The empirical performance of the method is assessed through simulation studies, where it is compared with state-of-the-art alternatives, demonstrating its competitive advantages. Finally, the practical potential of the approach is illustrated through an application to hourly NO$_2$ measurements over Lombardy, showing its ability to capture pollutant dynamics while accounting for sensor heterogeneity.

The remainder of the paper is organized as follows. Section \ref{Section:model} introduces the proposed physics-informed mixed-effects model for spatio-temporal data. Section \ref{Section:estimation} presents the estimation strategy for both parametric and nonparametric components. The asymptotic properties of the fixed and random effect estimators are discussed in Section \ref{Section:inference}. Section \ref{Section:test} evaluates the performance of the proposed method through simulation studies against state-of-the-art alternatives. The methodology is applied to air quality data in Section \ref{Section:application}, illustrating its ability to handle sensor heterogeneity in NO$_2$ assessment over Lombardy. Finally, Section \ref{Section:conclusions} summarizes the main contributions and outlines directions for future research.

\section{Physics-informed mixed-effect model for space-time data}
\label{Section:model}

Let $\{\mathbf{p}_i\}_{i=1}^n$ be a set of $n$ spatial locations on a bounded domain $\mathscr{D} \subset \mathbb{R}^2$, and let $\{t_j\}_{j=1}^m$ be a set of $m$ time instants in the interval $[0,T] \subset \mathbb{R}$. At these spatio-temporal locations, we partially observe $y_{ij}$, for $i=1,\dots,n, \ j=1,\dots,m.$  Assume that these observations are divided into $g$ non-overlapping groups, indexed by $k=1,\dots,g$. For each group $k$, we define the set of indices corresponding to the observed spatio-temporal locations as
\begin{equation*}
    \mathcal{O}_k = \{(i,j) \in \{(1,1),\dots,(n,m)\} \ \text{: the observation } y_{ij} \text{ belongs to group } k\}.
\end{equation*}
We denote by $|\mathcal{O}_k|$ its cardinality, and set $\mathcal{O} = \bigcup_{k=1}^g \mathcal{O}_k$.  For each pair $(i,j) \in \mathcal{O}$, we further observe $\mathbf{x}_{ij} \in \mathbb{R}^q$ and $\mathbf{z}_{ij} \in \mathbb{R}^p$, representing space-time varying fixed effect and random effect design vectors, respectively. Note that the definition of $\mathcal{O}$ naturally accounts for missing data, with $\cardO \le nm$.

For each group $k = 1, \dots, g$, we describe the observations through the following semiparametric mixed-effects model
\begin{equation}
\label{eq:model_k}
    \begin{split}
        & \yv_{k} = X_{k} \betav + \boldsymbol{f}_k + Z_{k} \bv_k + \varepsilonv_{k}, \\
        & \bv_k \overset{iid}\sim \mathcal{N}(\zerov, \Sigma_\mathbf{b}), \quad \varepsilonv_{k} \sim \mathcal{N}(\zerov,\sigma^2 I),
    \end{split}
\end{equation}
where $\yv_k$ is the vector collecting $\{y_{ij}\}_{(i,j)\in\mathcal{O}_k}$; $X_{k} \in \mathbb{R}^{|\mathcal{O}_k| \times q}$ stores, by rows, the covariates $\{\mathbf{x}_{ij}\}_{(i,j)\in\mathcal{O}_k}$, which affect the response $\yv_{k}$ through the fixed effects $\betav \in \mathbb{R}^q$, common to all observations; $Z_k \in \mathbb{R}^{|\mathcal{O}_k| \times p}$ collects the group-specific random effect design vectors $\{\mathbf{z}_{ij}\}_{(i,j)\in\mathcal{O}_k}$ associated with the random effects $\bv_k \in \mathbb{R}^p$; $\Sigma_\mathbf{b}$ represents the unknown covariance matrix of the random effects $\bv_k$; $\boldsymbol{f}_k$ is the vector of evaluations $f(\mathbf{p}_i, t_j)$, for $(i,j)\in \mathcal{O}_k$, representing a smooth function common to all groups and defined over the spatio-temporal domain $\mathscr{D} \times [0,T]$; finally, $\varepsilonv_k$ is the homoscedastic Gaussian within-group error for the $k$-th group, with variance $\sigma^2 > 0$. We assume that the fixed effect covariate matrix 
$X= \begin{bmatrix}
        X_1, 
        \cdots, 
        X_{g}
\end{bmatrix}^\top$ 
is full rank, and that the constant vector $\mathbf{1}$ does not belong to span$(X)$, to ensure ensure that the model is identifiable. Moreover, we assume that the random effects $\{\bv_k\}_{k=1}^g$ are independent of the error terms $\{\varepsilonv_k\}_{k=1}^g$, as standard in classical mixed-effects models. The validity of these assumptions can be assessed using well-established diagnostic tools (see, e.g., \cite{pinheiro_SandSplus}). The model in equation \eqref{eq:model_k} extends the one-level linear mixed-effects formulation in, e.g., \cite{pinheiro_SandSplus}, by including a nonparametric component $f$, in analogy with models introduced, e.g., in \cite{wood_Introduction_R}. The nonparametric term, as detailed in Section \ref{Subsection:physics}, captures the spatio-temporal structure of the data while incorporating prior physical information about the underlying phenomenon.

In the semiparametric model \eqref{eq:model_k}, we aim to estimate the smooth function $f$, the fixed effects $\betav$, and the random effect covariance matrix $\Sigma_\mathbf{b}$, which characterizes the variability across groups. To this end, leveraging the framework proposed in \cite{Arnone2023}, we minimize the following loss functional:
\begin{equation}
\label{eq:functional}
    J(\betav, f, \Sigma_\mathbf{b}) = \frac{1}{\cardO}
    \sum_{k=1}^g \sum_{(i,j)\in\mathcal{O}_k}
    \left( y_{ij} - \mathbf{x}_{ij}^\top \betav - f(\mathbf{p}_i, t_j) - \mathbf{z}_{ij}^\top \bv_k \right)^2 + \mathcal{P}_{\lambda_\mathscr{D}, \lambda_T}(f),
\end{equation}
The first term represents a quadratic data fidelity term, as in classical linear mixed-effects regression models, while the second term,  
$\mathcal{P}_{\lambda_\mathscr{D}, \lambda_T}(f)$, is a physics-informed penalty that enforces spatio-temporal smoothness of $f$ and incorporates prior knowledge on the physical dynamics of the process, as detailed in Section \ref{Subsection:physics}. The penalty  $\mathcal{P}_{\lambda_\mathscr{D}, \lambda_T}(f)$ is weighted by two positive parameters $\lambda_\mathscr{D}$ and $\lambda_T$, which control the trade-off between data fidelity, measured by the quadratic loss in the first term of \eqref{eq:functional}, and the spatial and temporal regularity of the field $f$, enforced by $\mathcal{P}_{\lambda_\mathscr{D}, \lambda_T}(f)$. The values of these smoothing parameters are selected using the Generalized Cross-Validation criterion \citep[see, e.g.,][]{craven1978smoothing, wahba1985comparison, Arnone2023}. 
Importantly, the functional \eqref{eq:functional} depends on $\Sigma_\mathbf{b}$ through the random coefficients $\bv_k$. This induces a non-quadratic structure of the estimation functional, which prevents a closed-form solution and necessitates the use of an iterative estimation scheme, as described in Section \ref{Section:estimation}.

\subsection{Including physical information}
\label{Subsection:physics}

In many applications, especially in environmental settings, prior physical knowledge about the phenomenon under study is available and can be incorporated into the model through suitable regularization terms. Such information is often formalized via partial differential equations (PDEs) describing the spatio-temporal evolution of the process. Here, we incorporate such PDE-based physical information into the statistical model through the penalty term
\begin{equation*}
\mathcal{P}_{\lambda_\mathscr{D}, \lambda_T}(f) 
= \lambda_\mathscr{D} \mathcal{P}_\mathscr{D}(f) + \lambda_T \mathcal{P}_T(f),
\end{equation*}
where
\begin{equation}
\label{eq:PDE_penalty}
\begin{aligned}
\mathcal{P}_\mathscr{D}(f) 
&= \int_{0}^{T} \int_{\mathscr{D}} \left( \mathcal{L}(\mathbf{p}) f(\mathbf{p},t) - u(\mathbf{p},t)\right)^2 
\, d\mathbf{p}\, dt, \\
\mathcal{P}_T(f) 
&= \int_{0}^{T}  \int_{\mathscr{D}} \left( \frac{\partial^2 f(\mathbf{p},t)}{\partial t^2} \right)^2 
\, d\mathbf{p} \, dt.
\end{aligned}
\end{equation}
The operator $\mathcal{L}(\mathbf{p})$ is a linear, second-order elliptic operator defining the diffusion-advection-reaction \gls{pde} $\mathcal{L}(\mathbf{p}) f(\mathbf{p},t) = u(\mathbf{p},t)$, where 
\begin{equation}
\label{eq:operator_L}
    \mathcal{L}(\mathbf{p}) f(\mathbf{p},t) = - \nabla \cdot \left(K(\mathbf{p}) \nabla f(\mathbf{p},t)\right) + \boldsymbol{\gamma}(\mathbf{p}) \cdot \nabla f(\mathbf{p},t) + c(\mathbf{p}) f(\mathbf{p},t).
\end{equation}
Specifically, the operator involves the gradient $\nabla = (\partial/{\partial p_1}, \partial / {\partial p_2})^\top$; the diffusion tensor $K: \mathscr{D} \rightarrow \mathcal{S}^+$ encoding (possibly anisotropic) diffusion processes, where $\mathcal{S}^+$ denotes the space of symmetric positive definite matrices; the advection field $\boldsymbol{\gamma}: \mathscr{D} \rightarrow \mathbb{R}^2$, which models directional transport in the spatial domain $\mathscr{D}$; and the scalar reaction term $c: \mathscr{D} \rightarrow \mathbb{R}$, which controls shrinkage of the field. The PDE parameters $K(\mathbf{p})$, $\boldsymbol{\gamma}(\mathbf{p})$, and $c(\mathbf{p})$ can vary over space, modeling different forms of physically-informed non-stationarity. Finally, the space-time varying forcing term $u: \mathscr{D} \times [0,T] \rightarrow \mathbb{R}$ in the spatial penalty $\mathcal{P}_\mathscr{D}(f)$ represents possible exogenous inputs. In this work, for simplicity, we restrict to the homogeneous case, with $u \equiv 0$, and we refer the reader to \cite{azzimonti_2014} for an analysis of nonhomogeneous forcing terms. 

The simplest instance of \eqref{eq:PDE_penalty} arises when $\mathcal{L}$ reduces to the Laplace operator, that is $(K, \boldsymbol{\gamma}, c) = (I, \zerov, 0)$, as considered by \cite{bernardi_modeling_2018} in the simpler setting of spatio-temporal regression with fixed effects only.  More general formulations, involving advection-diffusion dynamics, have been proposed in spatial regression without random effects by \cite{azzimonti_2014, azzimonti_blood_2015, arnone2019modeling} and, for the linear case, by \cite{tomasetto_2024}, as well as by \cite{des2025exploring} and \cite{castiglione_2025} for quantile regression problems. 

In general, $K$, $\boldsymbol{\gamma}$, and $c$ may depend on hyperparameters $\boldsymbol{\xi}$, which can be estimated from data using a parameter cascading algorithm \citep[see, e.g.,][]{ramsay_parameter_2007, xun_parameter_2013}, as discussed in \cite{bernardi_modeling_2018} and \cite{tomasetto_2024}. For instance, in the simulation studies in Section \ref{Section:test}, we model spatial anisotropy through a diffusion tensor $K(\boldsymbol{\xi})$, where $\boldsymbol{\xi}$ are unknown parameters controlling the direction and intensity of the anisotropy, estimated from the data via a parameter cascading approach; see \cite{bernardi_modeling_2018}. In the NO$_2$ case study over Lombardy, we instead adopt a diffusion-transport regularization term,
\[
-\Delta f(\mathbf{p}, t) + \xi \, \boldsymbol{\gamma}(\mathbf{p}) \cdot \nabla f(\mathbf{p}, t),
\]
which combines an isotropic diffusion component, representing atmospheric dispersion, with a space-varying transport term driven by the wind field $\boldsymbol{\gamma}$, encoding both its local direction and intensity. The relative contribution of the two components is governed by an unknown parameter, which is estimated from the data via parameter cascading; see \cite{tomasetto_2024} for details.

\subsection{Estimation problem}
\label{Subsection:optimization}
In order to properly define the estimation problem, we introduce suitable functional spaces for the nonparametric component $f$. Let $H^2(\mathscr{D})$ denote the second-order Sobolev space, i.e., the space of functions in $L^2(\mathscr{D})$ with square-integrable weak derivatives up to order two. We further denote by $L^2(0, T; H^2(\mathscr{D}))$ the space of square-integrable functions on $(0,T)$ with values in $H^2(\mathscr{D})$, and by $L^2(0, T; L^2(\mathscr{D}))$ the space of square-integrable functions on $(0,T)$ with values in $L^2(\mathscr{D})$.

The regularization terms in \eqref{eq:PDE_penalty} are well-defined on the space
\begin{equation*}
    V = \left \{ f \in L^2(0, T; H^2(\mathscr{D})) : \frac{\partial^2 f}{\partial t^2} \in L^2(0, T; L^2(\mathscr{D})) \right \}.
\end{equation*}
Let $\mathbf{\nu}$ denote the outward unit normal vector to the boundary $\partial \mathscr{D}$. We restrict $V$ by imposing homogeneous Neumann boundary conditions on $f$, namely $\nabla f \cdot \mathbf{\nu} = 0$, leading to the space
\begin{equation*}
    V_{BC} = \left \{ f \in V : \nabla f \cdot \mathbf{\nu} = 0 \ \text{on} \ \partial \mathscr{D} \times (0, T] \right \}.
\end{equation*}
Homogeneous Neumann boundary conditions are so-called natural conditions for the class of differential models considered here, and $V_{BC}$ provides a suitable functional embedding for the estimation of the nonparametric component $f$. Alternative boundary conditions, including Dirichlet, Robin, or mixed types, can also be considered, allowing for flexible modeling of the behavior of $f$ at the boundary of the domain $\mathscr{D}$; see, e.g., \cite{azzimonti_blood_2015} for details.

We thus consider the estimation problem
\begin{equation}
\label{eq:optimization}
     \underset{\betav \in \mathbb{R}^q, \ f \in V_{BC}, \ \Sigma_\mathbf{b} \in \mathcal{S}^+}{\mathrm{\argmin}} \, J(\betav, f, \Sigma_\mathbf{b}).
\end{equation}
We point out that, unlike simpler regression models considered, e.g., in \cite{bernardi2017penalized, arnone2019modeling, augustin_spacetime_2013, marra_modelling_2012, Arnone2023}, problem \eqref{eq:optimization} is not quadratic due to the presence of the covariance matrix $\Sigma_\mathbf{b}$. This makes the estimation problem more involved, requiring iterative strategies to approximate \eqref{eq:optimization}, as described in Section \ref{Section:estimation}.

\section{Model estimation}
\label{Section:estimation}

In this section, we present the iterative strategy used to solve the estimation problem in \eqref{eq:optimization} and to estimate $(\betav, f, \Sigma_\mathbf{b})$. To this end, we first reformulate model \eqref{eq:model_k} as a fixed effects regression model with correlated errors, following the classical approach described in \cite{wood_Introduction_R}. By collecting the group-specific quantities into block vectors, model \eqref{eq:model_k} can be written as
\begin{align*}
    &
    \underbrace{
    \begin{bmatrix}
        \yv_1 \\
        \yv_2 \\
        \vdots \\
        \yv_{g}
    \end{bmatrix}
    }_{\yv}
    =
    \underbrace{
    \begin{bmatrix}
        X_1 \\
        X_2 \\
        \vdots \\
        X_{g}
    \end{bmatrix}
    }_{X}
    \betav
    +
    \underbrace{
    \begin{bmatrix}
        \fv_1 \\
        \fv_2 \\
        \vdots \\
        \fv_{g}
    \end{bmatrix}
    }_{\fv} 
    + 
    \underbrace{
    \begin{bmatrix}
        Z_1 \bv_1 + \varepsilonv_1 \\
        Z_2 \bv_2 + \varepsilonv_2 \\
        \vdots \\
        Z_g \bv_{g} + \varepsilonv_g
    \end{bmatrix}
    }_{\ev},
\end{align*}
which can be written in compact form as
\begin{equation}
\label{eq:model_e}
  \yv = X \betav + \fv + \ev.
\end{equation}

The vector $\ev$ collects both the random effect contributions and the measurement errors, and  has a non-diagonal covariance structure. The formulation in \eqref{eq:model_e} shows that the proposed mixed-effects model can be equivalently expressed as a fixed effects regression model with heteroskedastic and correlated errors, where the random components $Z_k \bv_k$ and the noise terms $\varepsilonv_k$ are combined into a single stochastic term $\ev$.

Since, by assumption, $\bv_k$ and $\varepsilonv_k$ are independent Gaussian vectors for each $k$, it follows that
\[
\ev \sim \mathcal{N}(\zerov, \sigma^2 \Sigma_{\ev}),
\]
where $\Sigma_{e}$ is a block diagonal matrix with blocks $\Sigma_{e,k} = \frac{1}{\sigma^2} Z_k \Sigma_\mathbf{b} Z_k^\top + I_{\cardOk}$, for $k = 1, \dots, g$, and $I_{\cardOk}$ denotes the identity matrix of dimension $\cardOk$. The penalized negative log-likelihood of this model is then given by 
\begin{equation}
\label{eq:likelihood}
    \frac{1}{2\cardO}||(\sigma^2\Sigma_\mathbf{e})^{-1/2}(\yv - X \betav - \fv)||^2 + \frac{\log \! \left( \! \sqrt{(2\pi)^{\cardO} \sigma^2 \det(\Sigma_\mathbf{e})}\right)\!}{\cardO}  + \!\mathcal{P}_{\lambda_\mathscr{D}, \lambda_T}(f), 
\end{equation}
where $\det(\Sigma_\mathbf{e})$ denotes the determinant of $\Sigma_\mathbf{e}$.

Therefore, solving \eqref{eq:optimization} is equivalent to minimizing the penalized negative log-likelihood \eqref{eq:likelihood}. The resulting objective functional is non-quadratic, due to the dependence of $\Sigma_\mathbf{e}$ on the unknown covariance matrix $\Sigma_\mathbf{b}$, which induces a nonlinear coupling between the parameters and prevents a closed-form solution. This motivates the use of iterative estimation schemes. In particular, this formulation naturally leads to a \gls{fpirls} algorithm, extending the classical PIRLS method \citep{o1986automatic} to the functional setting. In the context of physics-informed penalized regression for spatially dependent data, related approaches have been proposed by \cite{wilhelm_2016} for generalized linear models and by \cite{castiglione_2025} for quantile regression. The estimation problem considered here is more involved, as it requires the joint estimation of the fixed effects $(\betav, f)$ and the variance-covariance matrix of the random effects $\Sigma_\mathbf{b}$. In particular, \eqref{eq:likelihood} cannot be minimized jointly with respect to $(\betav, f, \Sigma_\mathbf{e})$. We therefore adopt an iterative two-step procedure, as discussed in \cite{wood_Introduction_R} for the simpler purely parametric case. At each iteration, the algorithm alternates between the following two steps. In the first step, we estimate the fixed effects $(\betav, f)$ for a given covariance matrix $\Sigma_\mathbf{b}$, as detailed in Section \ref{Subsection:estimate_fixed-effects}. In the second step, we update the random effects covariance matrix $\Sigma_\mathbf{b}$ for given $(\betav, f)$, as described in Section \ref{Subsection:estimate_Sigmab}.

\subsection{Estimation of fixed effects}
\label{Subsection:estimate_fixed-effects}

We now describe the estimation procedure for the fixed effects $(\betav, f)$, for a given covariance matrix of the random effects $\Sigma_\mathbf{b}$. By neglecting the terms in \eqref{eq:likelihood} that do not depend on $(\betav, f)$, the problem reduces to the following penalized least-squares formulation:
\begin{equation}
\label{eq:optimization_fixed-effects}
    (\hat{\betav}, \hat{f}) = \underset{\betav \in \mathbb{R}^q, \ f \in V_{BC}}{\argmin} \left\{ \frac{1}{\cardO} \| \Sigma_\mathbf{e}^{-\frac{1}{2}} (\yv - X \betav - \fv) \|^2 + \mathcal{P}_{\lambda_\mathscr{D}, \lambda_T}(f)\right\}.
\end{equation}

Problem \eqref{eq:optimization_fixed-effects} can be solved using the techniques described in \cite{sangalli_spatial_2021}. In particular, to handle the PDE penalty and the possibly irregular geometry of the spatial domain $\mathscr{D}$, we adopt a numerical discretization scheme based on finite element bases in space and cubic B-spline bases in time. Let $\{\psi_1(\mathbf{p}),\dots,\psi_N(\mathbf{p})\}$ be a set of $N$ finite element basis functions (i.e., piecewise polynomial functions) defined on a triangulation $\mathscr{D}_\tau$ of the domain $\mathscr{D}$, and let $\{\varphi_1(t),\dots,\varphi_M(t)\}$ be a set of $M$ cubic B-spline basis functions defined on the time interval $[0,T]$.  Let $\Psi = \{\psi_\ell(\mathbf{p}_i)\}_{i,\ell} \in \mathbb{R}^{n \times N}$ be the matrix of spatial evaluations of the $N$ finite element bases at the $n$ locations $\{\mathbf{p}_1,\dots,\mathbf{p}_n\}$, and $\Phi = \{\varphi_r(t_j)\}_{j,r} \in \mathbb{R}^{m \times M}$ be the matrix of temporal evaluations of the $M$ spline bases at the $m$ time instants $\{t_1,\dots,t_m\}$.   We then represent the nonparametric term $f$ by the basis expansion
\[f(\mathbf{p}, t) = \sum_{\ell=1}^N \sum_{r=1}^M f_{\ell r}\,\psi_\ell(\mathbf{p})\,\varphi_r(t),
\]
and collect the coefficients of the basis expansion in the vector of coefficients $\mathbf{f} \in \mathbb{R}^{NM}$. 
To account for the missingness pattern of the data, we define $B$ as the sub-matrix of $\Phi \otimes \Psi$ obtained by removing the $(i+nj)$-th row whenever the datum at $(\mathbf{p}_i,t_j)$ is not observed. Additionally, we define the following matrices:
\begin{equation*}
    H = X (X^\top \Sigma_\mathbf{e}^{-1} X)^{-1} X^\top \Sigma_\mathbf{e}^{-1}, 
    \quad  
    Q = \Sigma_\mathbf{e}^{-1} (I - H).
\end{equation*}

Following the same arguments as in \cite{Arnone2023}, we can estimate the fixed effects estimator $(\check{\betav}, \check{\mathbf{f}}) \in \mathbb{R}^q \times \mathbb{R}^{NM}$, at each iteration of the FPIRLS algorithm, as:
\begin{equation}
\label{eq:closed_form_fixed-effects_estimator}
\begin{split}
    \check{\betav} &= \left(X^\top \Sigma_\mathbf{e}^{-1} X\right)^{-1} X^\top \Sigma_\mathbf{e}^{-1} (\yv - B\check{\mathbf{f}}), \\ 
    \check{\mathbf{f}} &= \frac{1}{\cardO} \left( \frac{1}{\cardO} B^\top Q B + P \right)^{-1} B^\top Q \yv, 
\end{split}
\end{equation}
where $P$ is the discrete counterpart of the penalty $\mathcal{P}_{\lambda_\mathscr{D}, \lambda_T}(f)$ in \eqref{eq:PDE_penalty}, as detailed in Section S1.5 of the supplementary material.

\subsection{Covariance estimation of random effects}
\label{Subsection:estimate_Sigmab}
In this section, we describe the procedure to estimate the covariance matrix of the random effects $\Sigma_\mathbf{b}$, for given $(\betav, f)$, obtained according to \eqref{eq:closed_form_fixed-effects_estimator}.

We define $D = \Sigma_\mathbf{b} / \sigma^2$ as the relative precision matrix of the model, and we denote by $\Delta$ a relative precision factor, that is, a matrix such that $D^{-1} = \Delta^\top \Delta$. Such a matrix $\Delta$ always exists, although it is not unique in general. We estimate $D$ via the Expectation-Maximization (EM) algorithm, which guarantees a monotone increase of the likelihood at each iteration. To this end, we introduce the following pseudo-data matrices:
\begin{equation*}
    \tilde\yv_k =
        \begin{bmatrix}
            \yv_k \\
            \zerov
        \end{bmatrix},
        \quad
        \tilde X_k =
        \begin{bmatrix}
            X_k \\
            \zerov
        \end{bmatrix},
        \quad
         \hat{\tilde{\fv_k}} =
         \begin{bmatrix}
            \hat{\fv_k} \\
            \zerov
         \end{bmatrix},
        \quad
        \tilde Z_k =
        \begin{bmatrix}
            Z_k \\
            \Delta
        \end{bmatrix},
        \quad \forall k = 1, \dots, g.
\end{equation*}
We denote by $R_k$ the upper triangular matrix in the QR decomposition of $\tilde Z_k$. We then define matrix $L \in  \mathbb{R}^{(g+1)p\times p}$ as
\begin{equation}
\label{eq:matrix_L}
    L = \begin{bmatrix}
             \hat{\bv}_1^\top / \sigma \\
            (R_1^{-1})^\top \\
            \vdots \\
            \hat{\bv}_g^\top / \sigma \\ 
            (R_g^{-1})^\top
    \end{bmatrix}.
\end{equation}
where $\hat{\bv}_k$ denotes the conditional maximum likelihood estimate of $\bv_k$, for $k = 1, \dots, g$. Finally, we denote by $A$ the triangular factor in the QR decomposition of $L$.

Within the EM algorithm, the E-step replaces the unobserved random effects $\bv_k$ with their conditional expectations, approximated here by $\hat{\bv}_k$. In the M-step, these estimates are plugged into the conditional likelihood of model \eqref{eq:model_e}, which is then maximized with respect to $D$ \citep{bates1998computational}.

The following proposition provides the analytical expression of the estimator of the relative precision matrix $D$.

\begin{proposition}
\label{prop:estimate_D}
For a given pair of fixed effects $(\hat{\betav}, \hat{f})$, the maximizer of the conditional likelihood of model \eqref{eq:model_e} is given by
\[
    \hat{D} = \frac{A A^\top}{g}.
\]
\end{proposition}

The proof is deferred to Section S1.1 of the supplementary material.

\section{Asymptotic distribution}
\label{Section:inference}

Consistently with the estimation strategy introduced in Section 3, we derive asymptotic results for the fixed effects, for a fixed value of $\Sigma_\mathbf{b}$.  Similarly, we study the asymptotic properties of the covariance estimator $\hat{\Sigma}_\mathbf{b}$ for fixed values of the fixed effect components.

Let $(\betaasynt, \fasynt) \in \mathbb{R}^{q} \times \mathbb{R}^{NM}$ denote the vector of discretized estimators of $(\betav, f)$ at FPIRLS convergence, where the notation $(\cdot)_{\cardO}$ indicates dependence on the sample size $\cardO$. We define the matrices
\begin{equation*}
    \Omega_{\cardO} = \cardO (B^\top Q B)^{-1}, 
    \quad 
    \Xi_{\cardO} = \frac{X^\top \Sigma_\mathbf{e}^{-1} X}{\cardO}.
\end{equation*}

The following two propositions state the asymptotic distributions of the fixed effect estimators $(\betaasynt, \fasynt)$.

\begin{proposition}
\label{prop:inference_f}
Assume that the limits
\[
\lim_{\cardO \to +\infty} \Omega_{\cardO} = \Omega, 
\quad 
\lim_{\cardO \to +\infty} \Xi_{\cardO} = \Xi,
\]
exist and are non-singular. If $\lambda_{\mathscr{D}} \sqrt{\cardO} \rightarrow \overline{\lambda}_\mathscr{D}$ and $\lambda_{T} \sqrt{\cardO} \rightarrow \overline{\lambda}_T$, for some finite values $\overline{\lambda}_\mathscr{D}$ and $\overline{\lambda}_T$, then $\fasynt$ has asymptotic distribution
\[
\sqrt{\cardO}(\fasynt - \mathbf{f}) \xrightarrow{\text{d}} \mathcal{N}_{NM}(\zerov, \sigma^2 \Omega).
\]
Moreover, $\fasynt$ is consistent for $\mathbf{f}$, that is, $\fasynt$ converges to $\mathbf{f}$ in probability.
\end{proposition}

\begin{proposition}
\label{prop:inference_beta}
Let $\{\fasynt\}$ be a sequence of consistent estimators for $\mathbf{f}$. Under the same assumptions stated in Proposition \ref{prop:inference_f}, the estimator $\betaasynt$ has asymptotic distribution
\[
\sqrt{\cardO} (\betaasynt - \betav) \xrightarrow{\text{d}} 
\mathcal{N}_q \left(\zerov, \ \sigma^2 \left( 
\Xi^{-1} + \frac{1}{\cardO^2} \Xi^{-1} (\Sigma_\mathbf{e}^{-1} X)^\top B \Omega B^\top (\Sigma_\mathbf{e}^{-1} X) \Xi^{-1} 
\right) \right).
\]
\end{proposition}

The following proposition provides the asymptotic distribution of the variance estimators $(\hat{\sigma}, \hat{\Sigma}_\mathbf{b})$ of the model, in the case of independent random effects.

\begin{proposition}
\label{prop:inference_Sigma_b}
Assume that the random effects $\bv_k$ are independent, that is, $\Sigma_\mathbf{b}$ is a diagonal matrix with diagonal $(\sigma^2_{b,1}, \dots, \sigma^2_{b,p})$. Then, for $\cardO \to +\infty$,
\begin{equation}
\label{eq:distr-random-effects}
    \begin{bmatrix}
        \log \hat{\sigma} \\
        \hat{\sigma}_{b,1} \\
        \vdots \\
        \hat{\sigma}_{b,p}
    \end{bmatrix}
    \xrightarrow{\text{d}}
    \mathcal{N} \left(
    \begin{bmatrix}
        \log \sigma \\
        \sigma_{b,1} \\
        \vdots \\
        \sigma_{b,p}
    \end{bmatrix}
    ,
    \mathcal{I}^{-1}_{\sigma, \Sigma_\mathbf{b}}
    \right),
\end{equation}
where $\mathcal{I}_{\sigma, \Sigma_\mathbf{b}}$ denotes the empirical information matrix, given by
\begin{equation*}
     \mathcal{I}_{\sigma, \Sigma_\mathbf{b}} 
    =
    \begin{bmatrix}
        \frac{2}{\hat{\sigma}^2} \sum_{k=1}^g \|\varepsilonv_k\|^2
        &
        0
        & 
        0 
        & 
        0
        \\
        0 
        &
        \frac{3}{\hat{\sigma}^4_{b,1}} \sum_{k=1}^g \|\bv_k\|^2 - \frac{g}{\hat{\sigma}^2_{b,1}}
        & 
        0
        & 0 
        \\
        \vdots
        & 
        \vdots
        &
        \vdots
        & 
        0
        \\
        0
        & 
        0
        &
        \cdots
        & 
        \frac{3}{\hat{\sigma}^4_{b,p}} \sum_{k=1}^g \|\bv_k\|^2 - \frac{g}{\hat{\sigma}^2_{b,p}}
    \end{bmatrix}.
\end{equation*}
\end{proposition}

The proofs of Propositions \ref{prop:inference_f}, \ref{prop:inference_beta}, and \ref{prop:inference_Sigma_b} are deferred to Sections S1.2, S1.3, and S1.4 of the supplementary material, respectively.
\section{Simulation study}
\label{Section:test}

In this section, we evaluate the performance of the proposed model, denoted as \gls{mestpde}, against state-of-the-art methods for spatio-temporal regression with mixed-effects. The proposed approach is implemented in the \texttt{fdaPDE} library \citep{fdapde_repo}.

We randomly sample $n=100$ spatial locations over the unit square domain $\mathscr{D} = [0,1]^2$, and partition them into $g=6$ groups. For the time dimension, we consider $m=11$ equispaced instants in the unit interval $[0,1]$. We consider the model
\begin{equation}
     \yv_{k} = X_{k} \betav + \fv_k + Z_k b_k + \varepsilonv_{k}, \quad k = 1, \dots, 6,
\end{equation}
where $Z_k = (1, \! \ldots , 1)^\top\!,$ that is, the model includes a random intercept only, as motivated by the NO$_2$ case study in Section \ref{Section:application}.  For the true spatio-temporal nonparametric term $f$, we consider a realization of a Gaussian field with Matérn covariance, generated using the \texttt{spate.sim} function from the R package \texttt{spate} \citep{spate}, specifying anisotropy with intensity $8$ and angle $\pi/4$.  For each time instant, we generate the two independent fixed effect covariates $X_k = (\mathbf{x}_{1,k}, \mathbf{x}_{2,k})$ as Gaussian random fields using the R function \texttt{grf} from the package \texttt{geoR} \citep{geoR}.  Further details about the generation of the fixed effect covariates, together with their spatio-temporal visualization, are reported in Section S2.1 of the Supplementary Material.  Finally, we set $\betav = (1,-1)$, $b_k \sim \mathcal{N}(0, \sigma_\mathbf{b}^2)$, and $\varepsilonv \sim \mathcal{N}(\zerov, \sigma^2 I_{\cardO})$, and we set $\sigma^2 = 1/4^2$ and $\sigma_\mathbf{b}$ so that the ratio $\sigma_\mathbf{b}^2 / (\sigma^2 + \sigma_\mathbf{b}^2)$ equals $0.30$. The generation of the noise $\varepsilonv$ is repeated $100$ times, obtaining $100$ independent replicas of the data.

In this setting, we consider the proposed \gls{mestpde} model, together with its isotropic variant, denoted as \texttt{MEST-ISO}, which assumes isotropic diffusion in the estimation process. Both \gls{mestpde} and \texttt{MEST-ISO} are implemented using a regular triangulation of the square domain with $476$ nodes and linear finite elements.  The \gls{mestpde} model includes a purely diffusive differential regularization term with unknown hyperparameters, introduced to capture anisotropy in the data. These hyperparameters, which determine the magnitude and orientation of the anisotropy, are estimated through the parameter cascading algorithm, as described in Section \ref{Section:model}. 

We compare these two methods with alternative techniques available in the literature, focusing on those implemented in widely used software. Specifically, we consider the Generalized Additive Mixed Model introduced in \cite{wood2006low} and implemented via the \texttt{gamm} function in the \texttt{mgcv} package \citep{mgcv}. As in the proposed \gls{mestpde} model, we adopt B-spline bases for the temporal discretization of all competing methods. For the spatial discretization, these approaches rely on thin plate spline bases \citep{Wahba1990SplineMF} and on soap film smoothing \citep{wood_soap_2008}. We refer to these two alternatives as \texttt{TPS} and \texttt{SOAP}, respectively.  We also consider the corresponding implementations based on the \texttt{gamm4} function \citep{gamm4}, which relies on the \texttt{lme4} framework \citep{lme4} instead of \texttt{nlme} \citep{nlme}, as in \texttt{gamm}. We denote these variants as \texttt{TPS4} and \texttt{SOAP4}, respectively.  For \texttt{TPS} and \texttt{TPS4}, we employ $75$ and $50$ spatial basis functions, respectively. For both \texttt{SOAP} and \texttt{SOAP4}, we use $8$ spatial basis functions, along with $8$ additional basis functions to define the boundary-interpolating soap film \citep[see][]{wood_soap_2008}. This configuration represents the maximum number of basis functions ensuring numerical stability, as larger values lead to execution failure.  

For the temporal dimension, all methods use $10$ cubic B-spline basis functions. Finally, smoothness selection is performed via Restricted Maximum Likelihood for \texttt{TPS4} and \texttt{SOAP4}, while for the other methods it is based on Generalized Cross-Validation. 

We have also explored the R package \texttt{sdmTMB} \citep[][]{anderson2022sdmtmb}, which is based on R-INLA \citep[][]{rue2009approximate}. However, this model produces unstable estimates of the nonparametric component $f$, which affect the overall results. For this reason, we exclude \texttt{sdmTMB} from the following discussion.

\begin{figure}[htbp]
\vspace{-3cm}
\centering
\resizebox{\textwidth}{!}{
\hspace{-1cm}\begin{tabular}{m{0.01\textwidth} m{.17\textwidth} m{.22\textwidth}  m{.20\textwidth} m{.15\textwidth} m{.15\textwidth}m{.15\textwidth}} 
    & \hspace{-1mm} \qquad $t=0$ & \quad $t=0.2$ & \quad \hspace{-1cm} $t=0.4$  &  \qquad \hspace{-1.9cm} $t=0.6$  & \qquad \hspace{-1.9cm} $t=0.8$  & \qquad \hspace{-1.5cm} $t=1$  \\
    \raisebox{0.5\height}{\rotatebox[origin=c]{90}{True}} & 
    \includegraphics[width=1.13\textwidth]{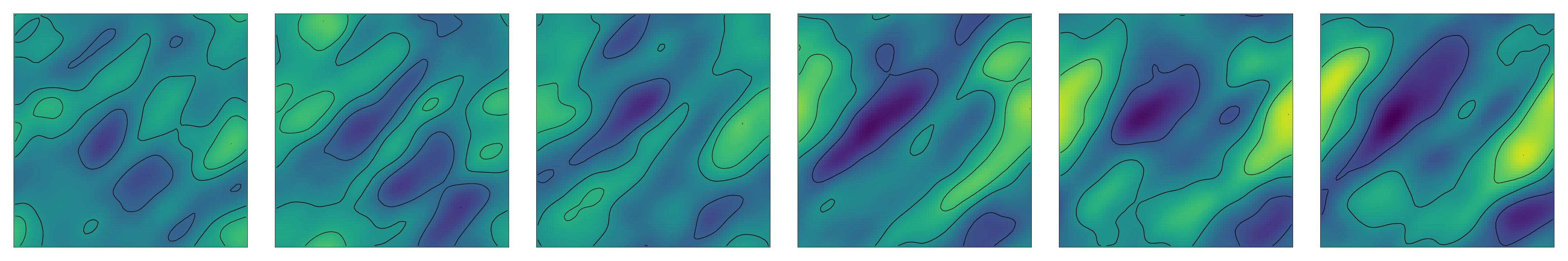} & \\
    \raisebox{0.5\height}{\rotatebox[origin=c]{90}{Data}} & 
    \includegraphics[width=1.13\textwidth]{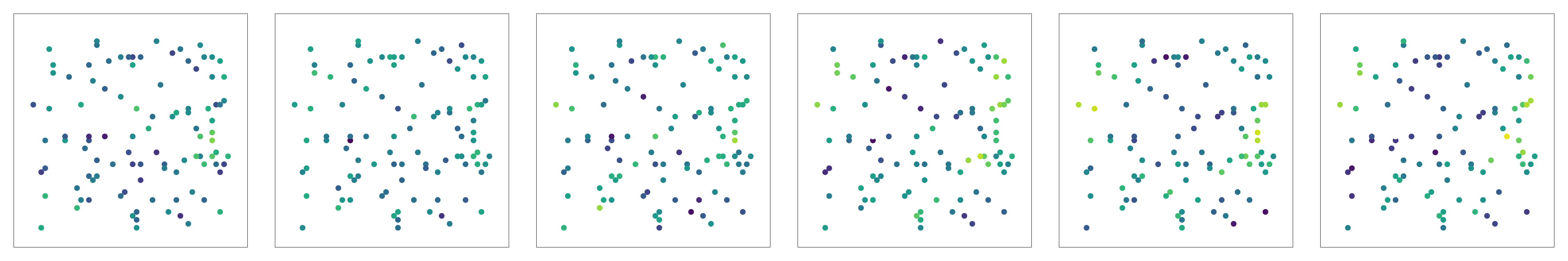} & \\  
    \raisebox{0.5\height}{\rotatebox[origin=c]{90}{\texttt{MEST-PDE}}} & 
    \includegraphics[width=1.13\textwidth]{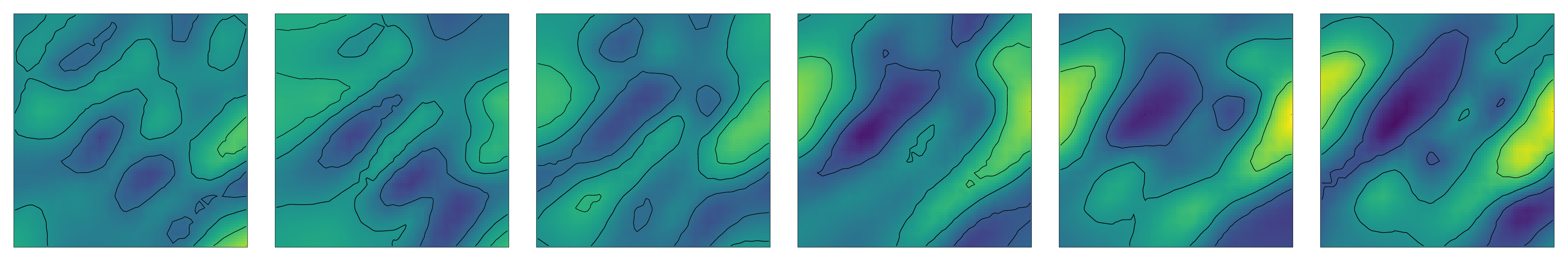} & \\
    \raisebox{0.5\height}{\rotatebox[origin=c]{90}{\texttt{MEST-ISO}}} & 
    \includegraphics[width=1.13\textwidth]{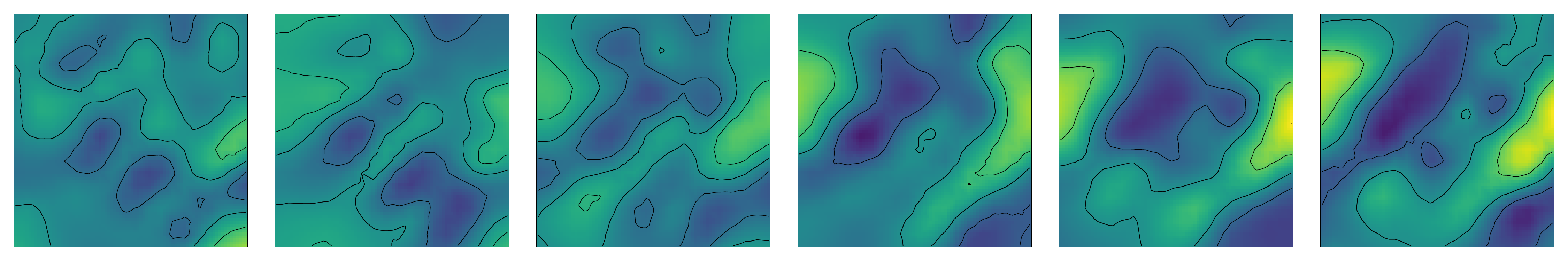} & \\
    \raisebox{0.5\height}{\rotatebox[origin=c]{90}{\texttt{TPS}}} & 
    \includegraphics[width=1.13\textwidth]{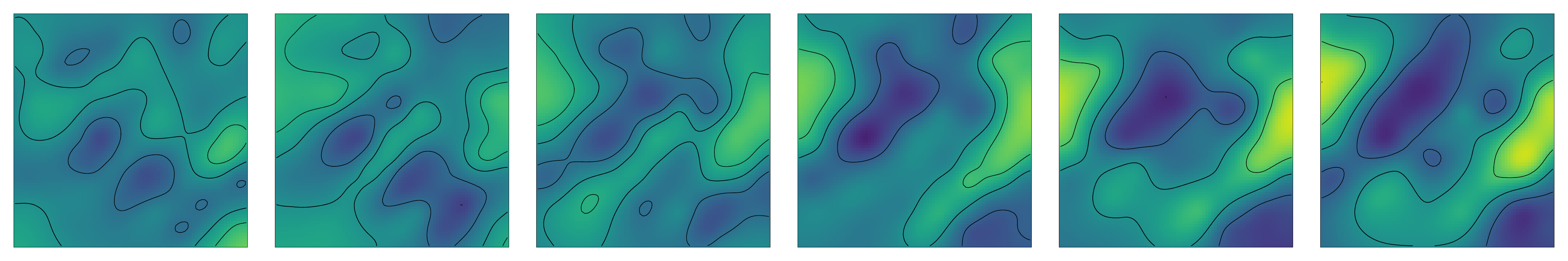} & \\
        \raisebox{0.5\height}{\rotatebox[origin=c]{90}{\texttt{TPS4}}} & 
    \includegraphics[width=1.13\textwidth]{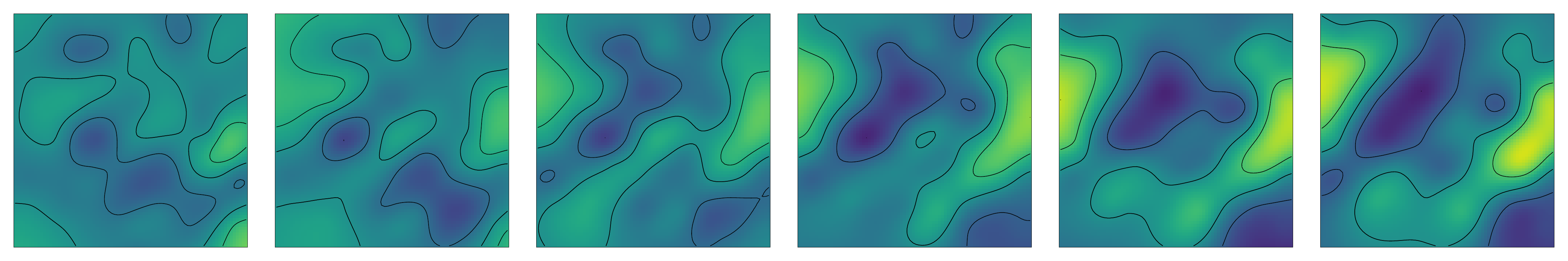} & \\
        \raisebox{0.5\height}{\rotatebox[origin=c]{90}{\texttt{SOAP}}} & 
    \includegraphics[width=1.13\textwidth]{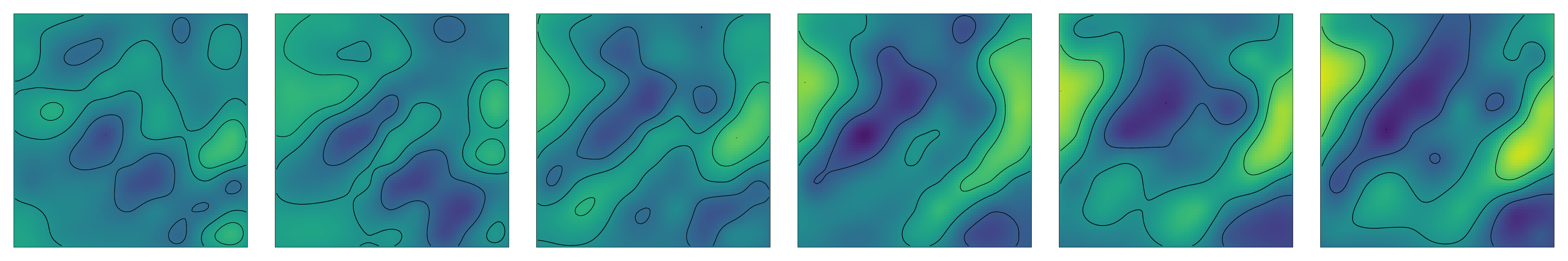} & \\
        \raisebox{0.5\height}{\rotatebox[origin=c]{90}{\texttt{SOAP4}}} & 
    \includegraphics[width=1.13\textwidth]{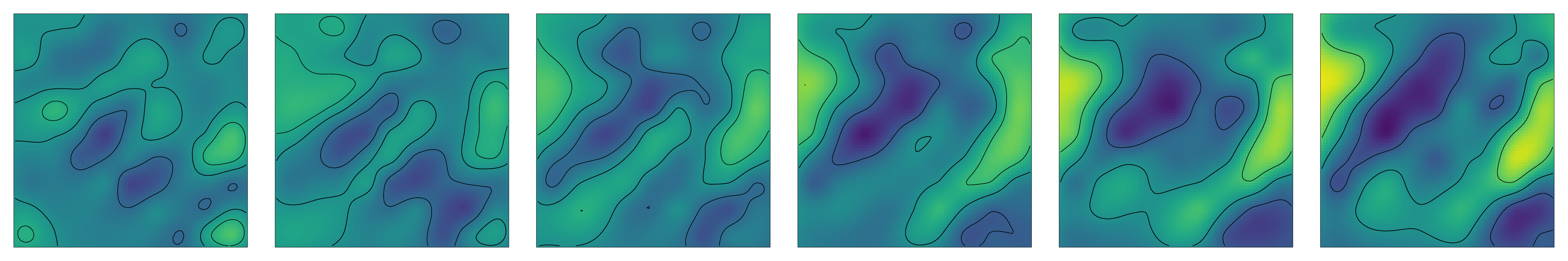} & \\

    \makebox[\textwidth][r]{\includegraphics[width=0.8\textwidth]{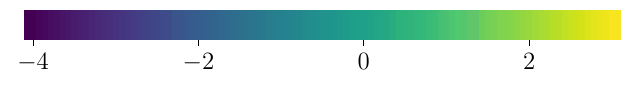}}

\end{tabular} 
}
\vspace{-0.5cm}
\caption{True and estimated nonparametric terms at selected time instants. The first row shows the true field, the second row displays the observed data for a fixed replica, while the subsequent rows report the estimated fields (averaged over the $100$ replicates) for each of the competing methods: the proposed Mixed-Effect Spatio-Temporal Regression with Partial Differential Equation regularization (\texttt{MEST-PDE}); its isotropic counterpart (\texttt{MEST-ISO}); Thin Plate Spline models based on \texttt{nlme} (\texttt{TPS}) and \texttt{lme4} (\texttt{TPS4}); and Soap film smoothing based on \texttt{nlme} (\texttt{SOAP}) and \texttt{lme4} (\texttt{SOAP4}).}
\label{fig:f}
\end{figure}

Model accuracy is assessed using the \gls{rmse} as a global measure of reconstruction error, which, for a given nonparametric term $f$, is defined as
\[
RMSE(\hat{f}) = \sqrt{\int_{\mathscr{D}}\int_0^T \left(f(\mathbf{p},t) - \hat{f}(\mathbf{p},t)\right)^2 \, d\mathbf{p} \, dt},
\]
where $\hat{f}$ represents the estimated field. The RMSE is approximated over a fine spatio-temporal grid of $27{,}500$ points.

\begin{figure}[htbp]
\centering
\begin{minipage}{0.32\textwidth}
    \centering
    \qquad \small{RMSE$(f)$} \\[0.3em]
    \includegraphics[width=\textwidth]{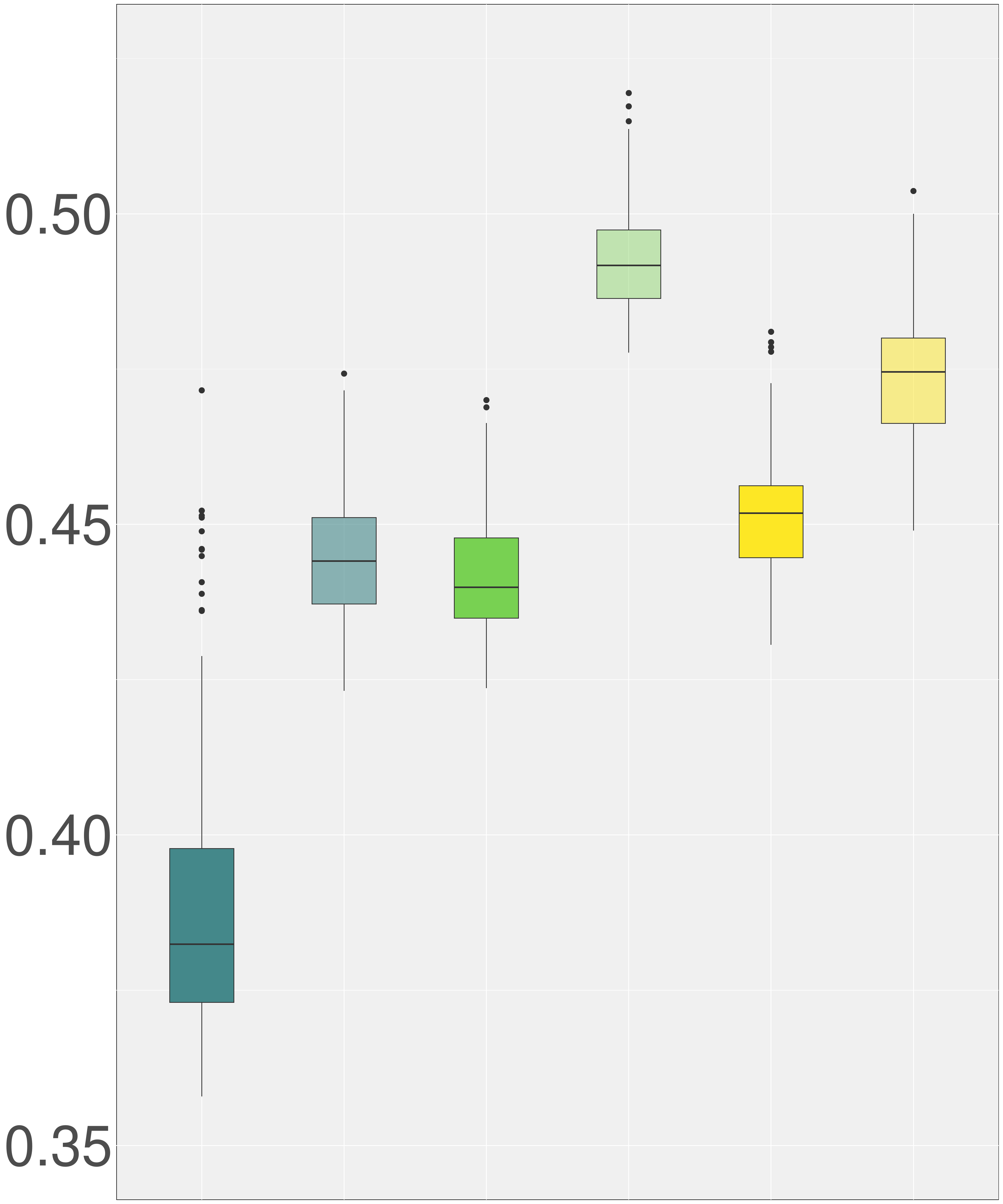}
\end{minipage} 
\begin{minipage}{0.32\textwidth}
    \centering
    \qquad \small{$\beta_1$}\\[0.3em]
    \includegraphics[width=\textwidth]{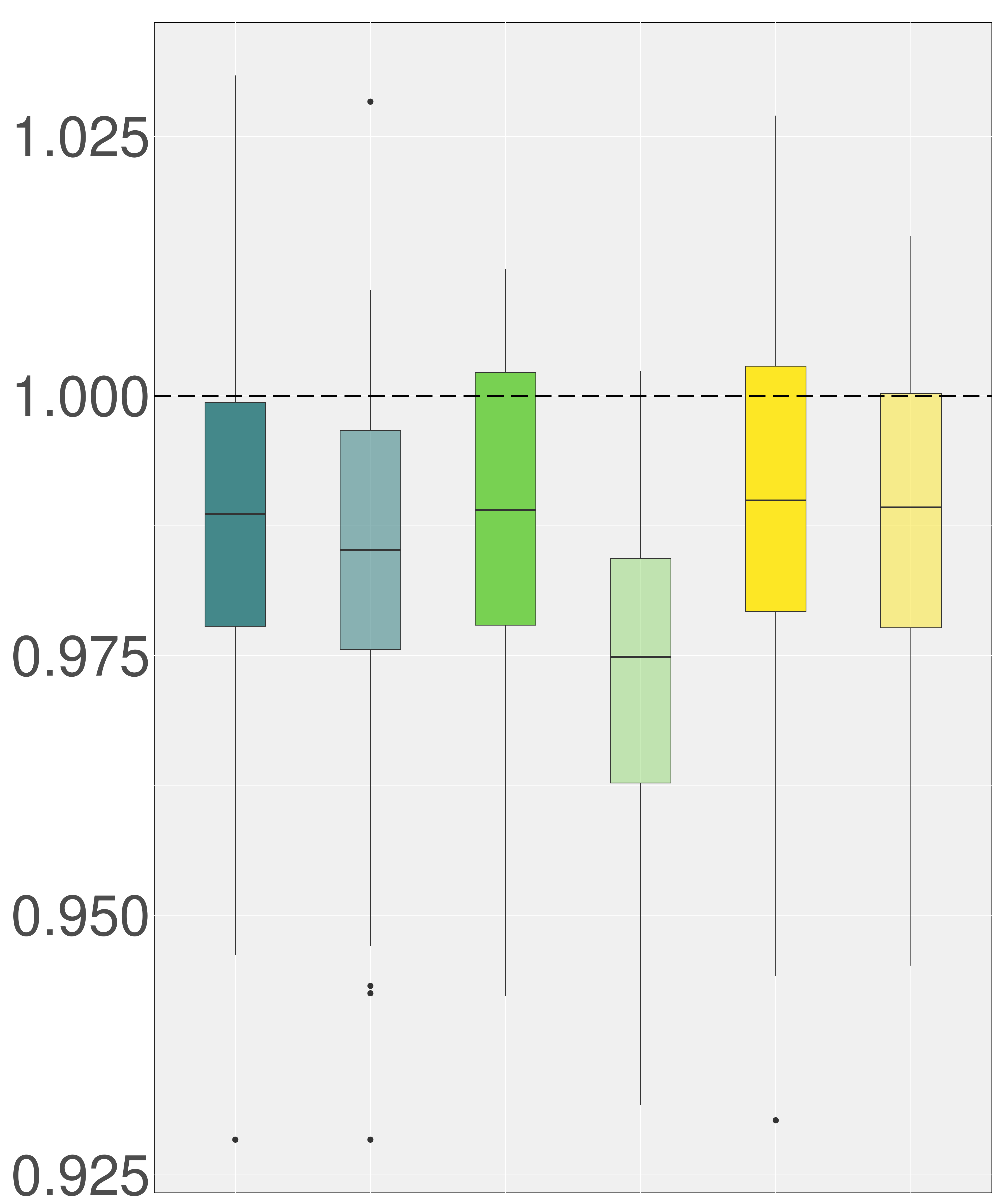}
\end{minipage}
\begin{minipage}{0.32\textwidth}
    \centering
    \qquad \small{$\beta_2$}\\[0.3em]
    \includegraphics[width=\textwidth]{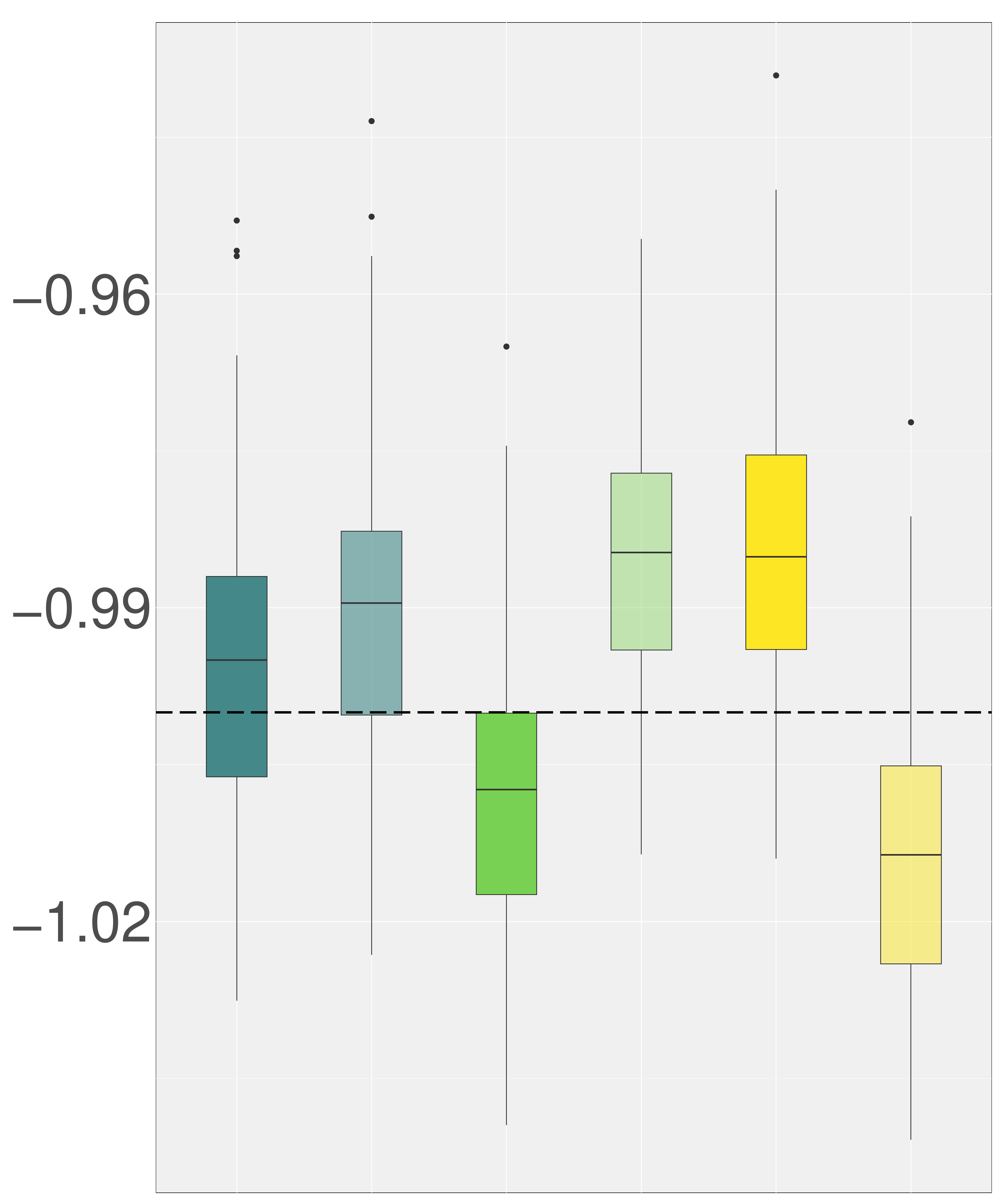} 
\end{minipage} 

\vspace{0.8em}

\includegraphics[width=0.8\textwidth]{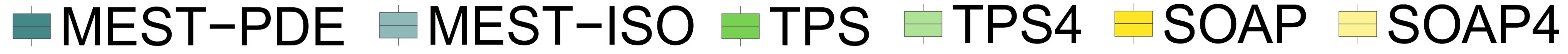}
\caption{Accuracy comparison of the fixed effect estimates obtained by the competing methods: the proposed Mixed-Effect Spatio-Temporal Regression with Partial Differential Equation regularization (\texttt{MEST-PDE}); its isotropic counterpart (\texttt{MEST-ISO}); thin plate spline models based on \texttt{nlme} (\texttt{TPS}) and \texttt{lme4} (\texttt{TPS4}); and soap film smoothing based on \texttt{nlme} (\texttt{SOAP}) and \texttt{lme4} (\texttt{SOAP4}). Left panel: RMSE of the nonparametric field $f$. Central panel: estimates of $\beta_1$. Right panel: estimates of $\beta_2$.}
\label{fig:fixed_effects}
\end{figure}

Figure \ref{fig:f} displays the observed data, together with the true and estimated nonparametric fields, at selected time instants. The \gls{mestpde} model more accurately captures the anisotropic features of the field, particularly at later time instants, where competing methods tend to oversmooth the central region of the domain.  This behavior is reflected in the RMSE of the nonparametric component, reported in the left panel of Figure \ref{fig:fixed_effects}, where \gls{mestpde} achieves the best performance, highlighting the benefit of incorporating anisotropic penalization. The last two panels of Figure \ref{fig:fixed_effects} show the estimates of the fixed effect regression coefficients. All methods exhibit comparable accuracy in estimating $\beta_1$, although a slight negative bias is observed, particularly for \texttt{TPS4}. In contrast, for $\beta_2$, the \gls{mestpde} model provides more accurate estimates than the competing approaches.

\begin{figure}[htbp]
\centering
\begin{minipage}{0.32\textwidth}
    \centering
    \small{$\sigma_\mathbf{b}$}\\[0.3em]
    \includegraphics[width=\textwidth]{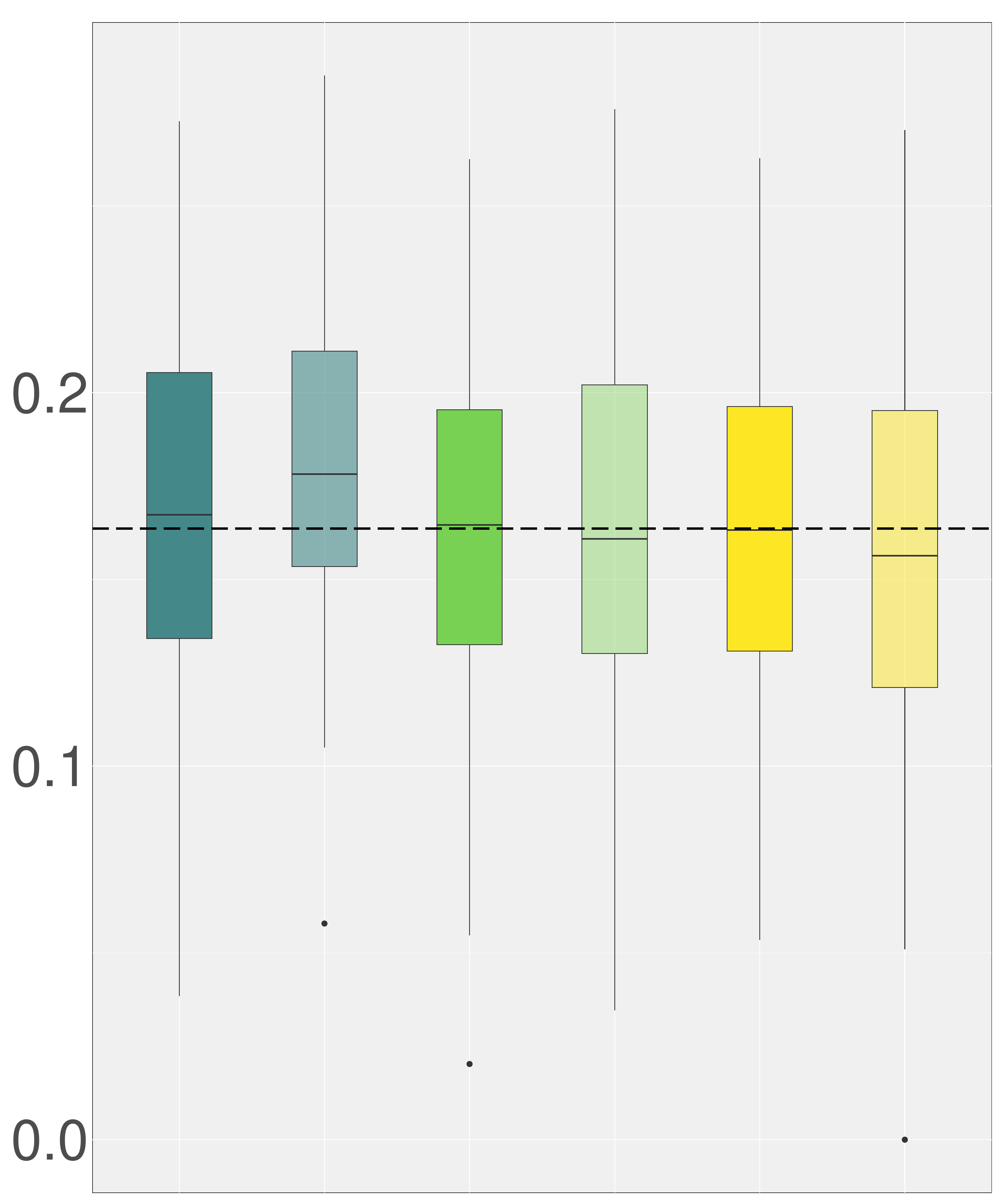}
\end{minipage} \hspace{0.5cm}
\begin{minipage}{0.32\textwidth}
    \centering
    \small{$D = \sigma_\mathbf{b}^2 / \sigma^2$}\\[0.3em]
    \includegraphics[width=\textwidth]{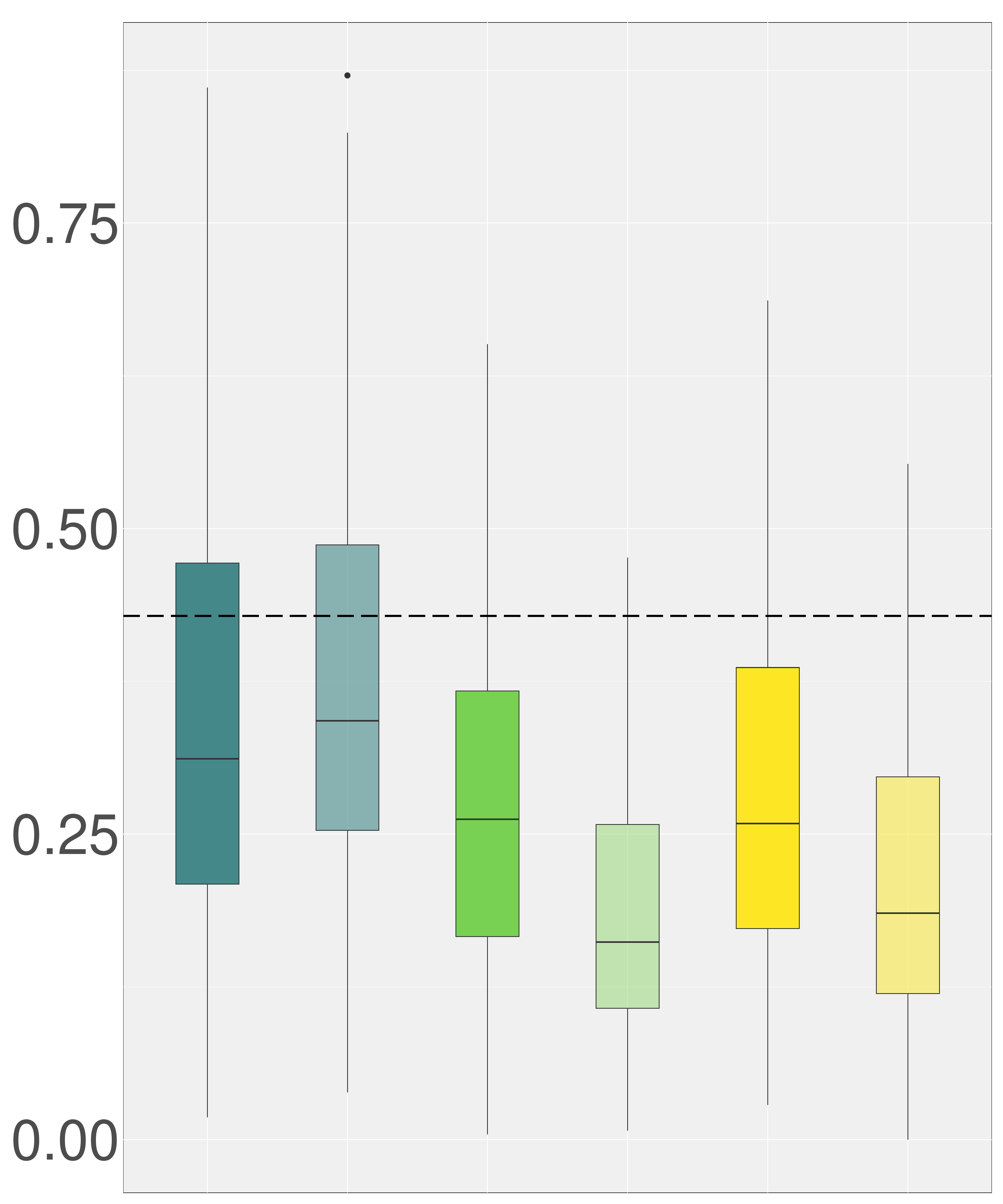}
\end{minipage}

\vspace{0.8em}

\includegraphics[width=0.8\textwidth]{img/simulations/legend_methods.jpg}
\caption{Estimated variance components obtained by the competing methods: the proposed Mixed-Effect Spatio-Temporal Regression with Partial Differential Equation regularization (\texttt{MEST-PDE}); its isotropic counterpart (\texttt{MEST-ISO}); thin plate spline models based on \texttt{nlme} (\texttt{TPS}) and \texttt{lme4} (\texttt{TPS4}); and soap film smoothing based on \texttt{nlme} (\texttt{SOAP}) and \texttt{lme4} (\texttt{SOAP4}). Left panel: estimated standard deviation of the random effects. Right panel: estimated relative precision factor.}
\label{fig:random_effects}
\end{figure}

Regarding the estimation of the random component, Figure \ref{fig:random_effects} shows the boxplots of the variance terms $\sigma_\mathbf{b}$ and $D$. All methods exhibit comparable performance in estimating $\sigma_\mathbf{b}$, with similar levels of variability.  For the precision factor $D$, the \gls{mestpde} and \texttt{MEST-ISO} models provide more accurate estimates, which can be attributed to a more reliable reconstruction of the noise variance $\sigma^2$.

Section S2 of the Supplementary Material reports additional simulation results comparing accuracy and computational cost under increasing sample sizes, highlighting the favorable scalability of the proposed MEST-PDE. These results show that the method is substantially faster than \texttt{TPS4} and \texttt{SOAP4}, while \texttt{TPS} and \texttt{SOAP} are computationally cheaper but significantly less accurate. In particular, increasing the number of spatial basis functions for \texttt{TPS} and \texttt{SOAP} is not feasible in practice, as it leads to execution failure, thereby preventing a comparison of computational costs at comparable levels of accuracy with \texttt{MEST-PDE}. Additionally, \texttt{MEST-PDE} shows a comparable or better asymptotic computational complexity than \texttt{TPS} and \texttt{SOAP}, and a significantly better asymptotic computational complexity than \texttt{TPS4} and \texttt{SOAP4}.

\section{Dealing with sensor heterogeneity in air quality assessment}
\label{Section:application}

In this section, we analyze the spatio-temporal concentration of nitrogen dioxide (NO$_2$) over Lombardy. This pollutant is primarily emitted from combustion processes, such as road traffic and industrial activities; elevated concentrations are associated with adverse effects on both human health and ecosystems \citep{chen2024long}. 
Within a single day, hourly NO$_2$ concentrations can vary substantially due to human-related factors such as traffic flows and heating demand. While these high-frequency measurements provide a detailed representation of air quality dynamics, they also pose additional modeling challenges due to stronger variability compared to daily or monthly aggregates. Nevertheless, capturing such short-term fluctuations is essential for exposure assessment, as even brief NO$_2$ peaks may have significant health impacts \citep[see, e.g.,][]{WHO2024}.
Figure \ref{fig:data} illustrates NO$_2$ measurements on $15$ January $2019$, provided by \gls{arpa} and publicly available from \cite{OpenData_lombardia}. The data exhibit a clear diurnal pattern, with higher concentrations in the morning and evening, reflecting traffic intensity and commuting behavior.
The dataset contains a small proportion of missing observations (approximately $2.41\%$ of the total), which, as discussed in the previous sections, can be naturally handled within our modeling framework. 
Additionally, analyzing the daily pattern of NO$_2$ concentrations allows us to incorporate information about the day’s wind field, which plays a crucial role in pollutant dispersion. In particular, the wind field, shown in Figure \ref{fig:wind}, is introduced in the regularizing term as an appropriate transport term, as described below.

As anticipated in Section \ref{Section:intro}, we consider distinct sources of variability in NO$_2$ concentrations, in order to disentangle the contributions of physical, morphological, anthropogenic, and technological factors influencing air pollution levels.
To account for the heterogeneity of measurement technologies across the \gls{arpa} monitoring network, we include in the statistical model a random effect, specified as a station-specific random intercept $b_k$. This term captures the variability introduced by the measurement process of the instrumentation, allowing the model to separate sensor-related noise from the underlying spatio-temporal dynamics of the phenomenon of interest. 
Note that a purely fixed effect model including dummy variables would not capture the variability induced by the measurement process.
The spatial distribution of the different sensor technologies, namely \textit{API}, \textit{enviro}, \textit{serinus}, and \textit{thermo}, is illustrated in the bottom-right panel of Figure \ref{fig:data}.

\begin{figure}[tbp]
\small{\hspace{0.06\textwidth} Lombardy map \hspace{0.17\textwidth}  Altitude \hspace{0.15\textwidth} Population density}\\
\includegraphics[width=\textwidth]{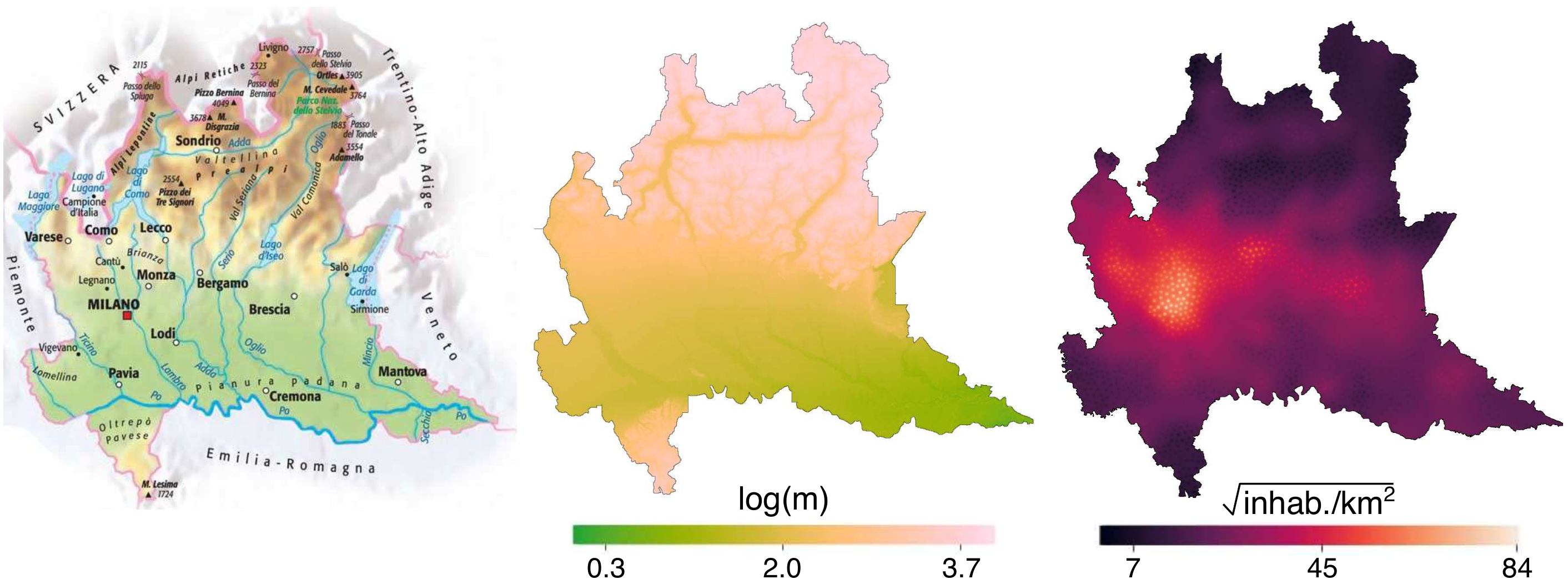}

\caption{Left panel: geographical map of the Lombardy region (source: \href{https://cartinadatieuropa.it/italia/lombardia/cartina/}{cartinadatieuropa.it}). Central panel: logarithm of the altitude data derived from the Digital Elevation Model. Right panel: square root of the population density computed from ISTAT census data.}
\label{fig:map_and_covariates}
\end{figure}

Another source of variability arises from both geographical and anthropogenic characteristics of the territory. According to recent literature, natural and human-driven factors are key drivers of NO$_2$ formation \citep[see, e.g.,][]{salama_satellite_2022, california_air_sources_board}.  From a morphological perspective, the Po Valley is subject to frequent thermal inversion phenomena, which limit air circulation and contribute to the persistence of pollutants in lowland areas \citep{trinh2019temperature}. To account for this effect, we include altitude as a fixed effect covariate, distinguishing between lowland zones, where thermal inversion tends to trap pollutants near the surface, and mountainous areas, where cleaner air and stronger circulation mitigate pollution levels. Elevation data are obtained from a Digital Elevation Model provided by the National Institute of Geophysics and Vulcanology \citep{INGV}, and are displayed in the central panel of Figure \ref{fig:map_and_covariates}.  Human activities such as traffic, residential heating, and industrial production also represent major sources of NO$_2$. These factors are closely related to population density, which is therefore included in the model as a second fixed effect covariate. Population density data, shown in the right panel of Figure \ref{fig:map_and_covariates}, are provided by the Istituto Nazionale di Statistica and sourced from \cite{Geoportale_lombardia_site}.

A third and final component we consider in modeling NO$_2$ concentrations is wind dynamics, which drive the transport of the pollutant. For this reason, we regularize the nonparametric component $f$ through an advection-diffusion \gls{pde}, which combines an isotropic diffusion term, modeling the dispersion of the pollutant in the air, with a transport term, accounting for advection driven by the wind field. The resulting spatial penalty is given by
\begin{align}
\label{eq:penalty_arpa}
    \mathcal{P}_{\mathscr{D}}(f) =  \int_{0}^{T} \int_{\mathscr{D}}  \big( -\nabla \cdot (I \nabla f(\mathbf{p}, t)) + \xi \ \boldsymbol{\gamma}(\mathbf{p}) \cdot \nabla f(\mathbf{p}, t) \big)^2 
    \, d\mathbf{p}\, dt,  
\end{align}
where $\boldsymbol{\gamma}$ is the space-varying daily average wind vector, shown in Figure \ref{fig:wind}, obtained from data at $119$ \gls{arpa} meteorological stations distributed across the region and collected on the same day as the NO$_2$ measurements. The unknown parameter $\xi$, which balances the relative contribution of the diffusion term $-\nabla \cdot (I \nabla f)$ and the transport term $\boldsymbol{\gamma} \cdot \nabla f$, is estimated from the data via a parameter cascading strategy, as described in \cite{tomasetto_2024}. It is important to note that the transport field $\boldsymbol{\gamma}(\mathbf{p})$ considered in the present framework is spatially varying but assumed to be time-invariant. This assumption is appropriate when analyzing intra-day dynamics, as the wind effect can be reasonably approximated by its daily average. However, it becomes restrictive when extending the analysis to multiple days, since wind conditions may vary substantially over time and cannot be consistently represented by a single transport field. Extending the model to incorporate time-varying transport dynamics is an active direction of ongoing research, as discussed in the final section.

\begin{figure}
\label{fig:mu_fields}
\hspace{-0.6cm}
\begin{tabular}{p{0.0\textwidth}p{.31\textwidth}p{.31\textwidth}p{.31\textwidth}}
& \quad \qquad \quad \small{08:00} &  \qquad \quad \quad \small{16:00} &  \quad \quad \qquad \small{21:00} \\
&\includegraphics[width=.31\textwidth]{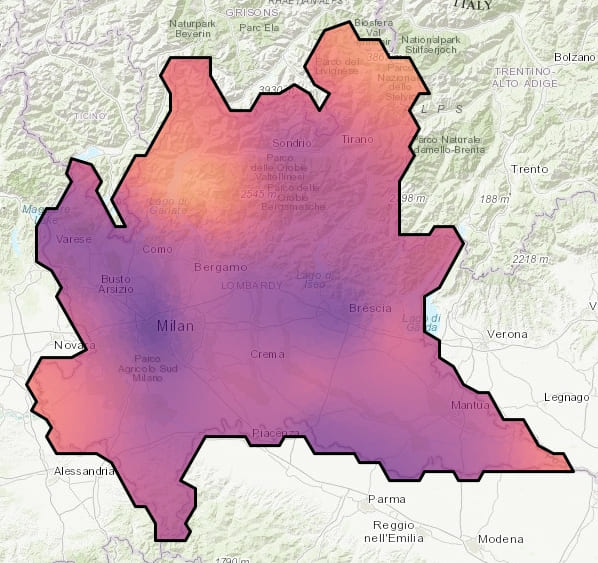}
&\includegraphics[width=.31\textwidth]{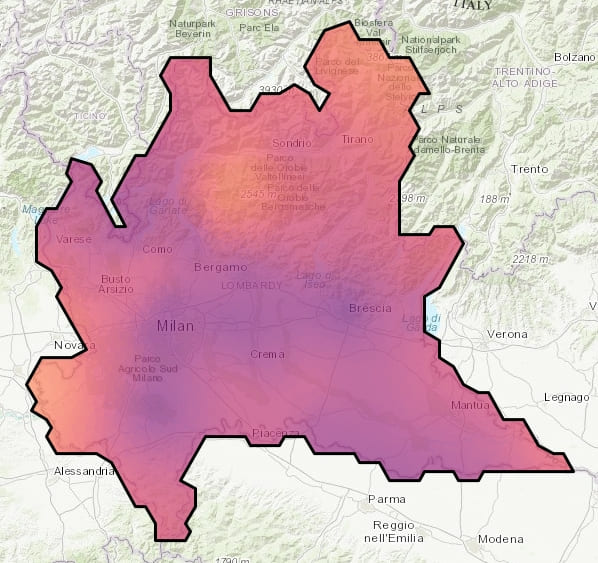}
&\includegraphics[width=.31\textwidth]{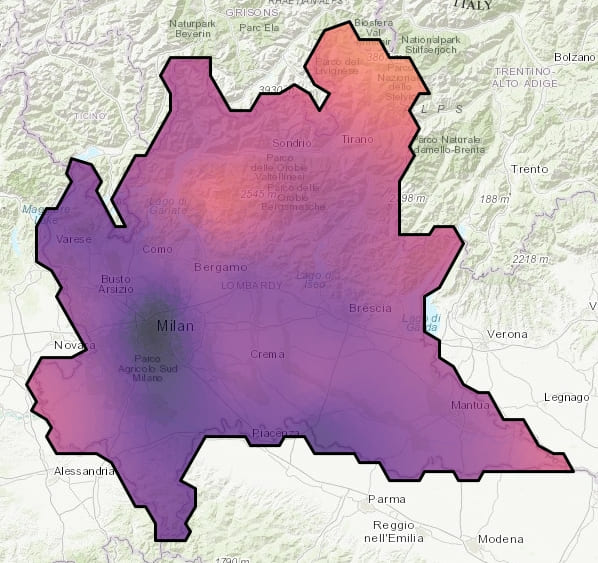}
\end{tabular}

\centering
\includegraphics[width=0.5\textwidth]{img/colorbars/field_arpa_colorbar.pdf}

\caption{Estimated spatial fields at three representative hours of the day: 08:00 (left panel), 16:00 (central panel), and 21:00 (right panel).}
\label{fig:fields}
\end{figure}

Using the mixed-effects model with the specifications described above, we estimate the spatio-temporal concentration levels of NO$_2$ over the Lombardy region. The results are shown in Figure \ref{fig:fields} for three representative hours of the day, namely 08:00, 16:00, and 21:00. 
The estimated maps highlight the typical diurnal cycle of NO$_2$. In the morning, concentrations increase markedly due to traffic and domestic activities, with pronounced peaks in major urban centers such as Milano and Brescia. In the afternoon (16:00), levels decrease slightly, reflecting reduced mobility during working hours and a temporary stabilization of emissions. Nevertheless, urban hotspots remain clearly visible, indicating the persistent impact of NO$_2$ in metropolitan areas. 
In the evening (21:00), elevated concentrations are observed across much of the Po Valley, as limited air circulation favors the accumulation of pollutants over the course of the day, particularly in lowland and urban areas, while mountainous regions remain less affected. These patterns are consistent with the expected behavior of NO$_2$ under typical urban and meteorological conditions.

Moreover, contrary to what would be expected under a regular diurnal cycle, NO$_2$ concentrations at 00:00 are substantially lower than those observed at the end of the day. This behavior can be attributed to strong wind conditions recorded in the days preceding the observations, namely the $13^{\text{th}}$ and $14^{\text{th}}$ of January. In particular, a weather alert was issued on the $13^{\text{th}}$, when wind gusts reached up to $70$ km/h, which is highly unusual for the region.  Finally, we observe that the area surrounding the city of Sondrio, together with the mountain towns in the northwestern part of Lombardy, consistently exhibits lower NO$_2$ concentrations throughout the day. This pattern can be explained by the Alpine morphology of the region. Indeed, low population density and limited anthropogenic activities reduce local emissions, while stronger wind circulation, characteristic of mountainous areas, enhances pollutant dispersion.

\begin{figure}[htbp]
    \small{\hspace{0.28\textwidth}Estimated hourly NO$_2$ \hspace{0.25\textwidth} Dotplot}\\
      \setlength{\unitlength}{1cm}
    \begin{picture}(0,0)
        \put(-0.1,2){\rotatebox{90}{\scriptsize  NO$_2$ [$\sqrt{\mu g/m^3}$]}}
    \end{picture}  
    \includegraphics[width=0.95\textwidth]{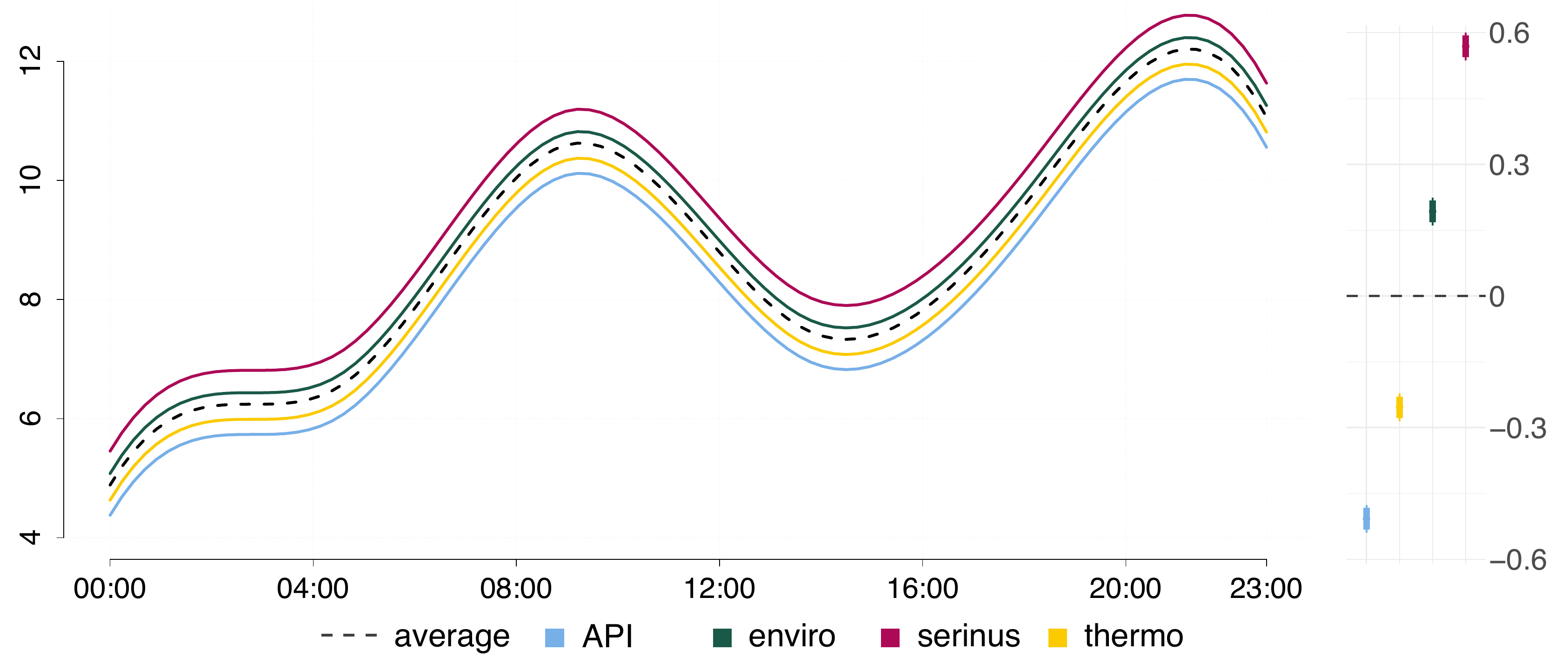}
\caption{Left panel: estimated hourly NO$_2$ time profile for the city of Milano (dashed black curve), together with the effect of the measurement-specific technologies on the average signal (solid colored curves). Right panel: estimated $99\%$ confidence bands for the measurement-specific random intercepts. The intervals are ordered according to the sign of the pointwise estimates $\hat{\bv}_k$.
}
\label{fig:estimated_time_profile}
\end{figure}

To illustrate the impact of the random effects on the estimated fields, we analyze the temporal evolution of NO$_2$ concentrations at a fixed location, corresponding to the city of Milano. Figure \ref{fig:estimated_time_profile} shows the estimated hourly average profile (dashed black line), together with the sensor-specific curves associated with the $4$ technological categories considered in this study (solid colored lines).   The average curve reproduces the main diurnal pattern, characterized by a pronounced morning peak, a moderate decline in the afternoon, and persistent pollutant levels in the evening, consistent with the spatial patterns previously observed. The sensor-specific curves highlight the effect of the measurement technology on the estimated signal.  The correction associated with the measurement technology, captured by the random component, is modest in magnitude, as expected for well-calibrated sensors. However, the $99\%$ confidence bands of the estimated $b_k$ in the right panel of Figure \ref{fig:estimated_time_profile} indicate that these sensor-specific effects, although small, are significantly different from zero. Modeling sensor technology as a random effect allows us to estimate the variance $\sigma_\mathbf{b}^2$ and directly quantify the portion of variability attributable to the measurement devices. This is summarized by the ratio $\text{PVRE} = \hat{\sigma}_\mathbf{b}^2 / (\hat{\sigma}_\mathbf{b}^2 + \hat{\sigma}^2) = 13.52\%$, confirming that the random component captures a non-negligible share of the total variability. In summary, the proposed model enables a clear separation between the spatio-temporal field, which reflects the underlying pollutant dynamics, and the random component, which accounts for systematic deviations due to sensor characteristics.

Section S3 reports a leave-one-station-out comparison of the proposed approach against \texttt{TPS} and \texttt{SOAP}, demonstrating the higher accuracy of \texttt{MEST-PDE}, for instance in terms of the Root Mean Squared Error.

\section{Discussion and further developments}
\label{Section:conclusions}

In this work, we have introduced a novel physics-informed semiparametric mixed-effects model for the analysis of spatio-temporal data. The proposed framework extends the PDE-regularized regression approach of \cite{sangalli_spatial_2013} and \cite{azzimonti_2014, azzimonti_blood_2015} by incorporating random effects into the statistical model. The inclusion of random effects broadens the applicability of the framework to settings with grouping structures, enabling a flexible representation of group-specific variability.

This work can be extended in several directions. First, one could consider introducing additional levels of random effects, extending the proposed framework to multilevel modeling in order to represent complex nested hierarchical structures, which are often encountered in real-world applications. Furthermore, extending the class of physics-informed penalties is also of interest. For instance, allowing for time-varying PDE parameters would enable the modeling of phenomena in which spatial anisotropy and stationarity evolve over time. In the case of NO$_2$ measurements discussed in Section  \ref{Section:application}, incorporating a time-varying wind field would allow the analysis to be extended to longer time periods. Such a generalization poses relevant theoretical, methodological, and computational challenges, which are currently under investigation. Another important direction concerns spatial confounding, which arises when covariates exhibit spatial autocorrelation. While recent work in semiparametric regression has addressed spatial confounding for fixed effect covariates \citep{dupont2022spatial}, analogous developments for mixed-effects semiparametric models of the type considered here are still lacking.

Additionally, from a computational perspective, the efficiency of the estimation procedure could be improved in the final stages of the optimization. Indeed, the EM algorithm typically converges rapidly in the initial iterations, but tends to slow down near convergence. A possible solution, proposed by \cite{bates1998computational}, consists of a hybrid strategy that combines EM iterations in the early phase of the algorithm with a Newton-type optimization in the final phase, thereby accelerating convergence without compromising stability.  Finally, the proposed framework naturally extends to more complex spatial domains, including Riemannian manifolds \citep[][]{ettinger2016spatial, lila2016smooth}, three-dimensional regions with complex geometries \citep[][]{arnone2023analyzing}, and graphs \citep[][]{clemente2026nonparametric}, where accounting for the domain geometry is essential for an accurate representation of spatial dependence.

\section*{Acknowledgments}
We are grateful to the Editor and Reviewers for their insightful comments and suggestions, which have significantly improved the quality and clarity of the manuscript. We thank Alessandro Melchionda for his preliminary work on this topic in his Master Thesis at Politecnico di Milano.
This work has been supported by the project GRINS - Growing Resilient, INclusive and Sustainable (GRINS PE00000018 – CUP D43C22003110001), funded by the European Union -  NextGenerationEU programme. The views and opinions expressed are solely those of the authors and do not necessarily reflect those of the European Union, nor can the European Union be held responsible for them. M.F.\ De Sanctis, F. \ Ieva and L.M.\ Sangalli acknowledge MUR research project Dipartimento di Eccellenza 2023 - 2027, Dipartimento di Matematica, Politecnico di Milano. 
The simulations discussed in this work were performed on the HPC Cluster of the Department of Mathematics of Politecnico di Milano which was funded by MUR grant Dipartimento di Eccellenza 2023-2027. L.M.\ Sangalli also acknowledges the PRIN 2022 project CoEnv - Complex Environmental Data and Modeling (CUP 2022E3RY23), funded by the European Union – NextGenerationEU programme, and by the Italian Ministry for University and Research.

\bibliographystyle{elsarticle-harv} 
\bibliography{bibliography}

@article{Arnone2023,
    author  = "Arnone, Eleonora and Sangalli, Laura M. and Vicini, Andrea",
    title   = "Smoothing spatio-temporal data with complex missing data patterns",
    journal = "Statistical Modelling",
    year    = "2023",
    volume  = "23",
    number  = "4",
    pages   = "327--356",
    doi     = "https://doi.org/10.1177/1471082X211057959"
}

@article{azzimonti_blood_2015,
  author = {Laura Azzimonti and Laura M. Sangalli and Piercesare Secchi and Marta Domanin and Fabio Nobile},
  title = {Blood Flow Velocity Field Estimation Via Spatial Regression With {PDE} Penalization},
  journal = {Journal of the American Statistical Association},
  volume = {110},
  pages = {1057--1071},
  year = {2015},
doi = {https://doi.org/10.1080/01621459.2014.946036}
}

@article{azzimonti_2014,
    AUTHOR = {Azzimonti, Laura and Nobile, Fabio and Sangalli, Laura M. and Secchi, Piercesare},
     TITLE = {Mixed finite elements for spatial regression with {PDE} penalization},
  JOURNAL = {SIAM/ASA Journal on Uncertainty Quantification},
    VOLUME = {2},
      YEAR = {2014},
    NUMBER = {1},
     PAGES = {305--335},
doi={http://dx.doi.org/10.1137/130925426}
}

@article{tomasetto_2024,
    AUTHOR = {Tomasetto, Matteo and Arnone, Eleonora and Sangalli, Laura M.},
     TITLE = {Modeling anisotropy and non-stationarity through physics-informed spatial regression},
   JOURNAL = {Environmetrics},
    VOLUME = {},
      YEAR = {2024},
    NUMBER = {},
     PAGES = {},
    doi = {https://doi.org/10.1002/env.2889}
}

@article{bernardi_modeling_2018,
    AUTHOR = {Bernardi, Mara S. and Carey, Michelle and Ramsay, James O. and Sangalli, Laura M.},
     TITLE = {Modeling spatial anisotropy via regression with partial differential regularization},
  JOURNAL = {Journal of Multivariate Analysis},
    VOLUME = {167},
      YEAR = {2018},
     PAGES = {15--30},
doi={https://doi.org/10.1016/j.jmva.2018.03.014}
}

@article{wilhelm_2016,
  author = {Martina Wilhelm and Laura M. Sangalli},
  title = {Generalized spatial regression with differential regularization},
  journal = {Journal of Statistical Computation and Simulation},
  volume = {86},
  pages = {2497--2518},
  year = {2016}, 
  doi = {https://doi.org/10.1080/00949655.2016.1182532}
}

@article{sangalli_spatial_2021,
  author = {Laura M. Sangalli},
  title = {Spatial Regression With Partial Differential Equation Regularisation},
  journal = {International Statistical Review},
  volume = {89},
  pages = {505--531},
  year = {2021},
 doi = {https://doi.org/10.1111/insr.12444}
}

@article{castiglione_2025,
  author = {Cristina Castiglione and Eleonora Arnone and Marco Bernardi and Alessio Farcomeni and Laura Sangalli},
  title = {{PDE}-regularised spatial quantile regression},
  journal = {Journal of Multivariate Analysis},
  volume = {205},
  year = {2025},
    doi = {https://doi.org/10.1016/j.jmva.2024.105381}
}

@article{bates1998computational,
  title={Computational methods for multilevel modelling},
  author={Bates, Douglas M and Pinheiro, Jos{\'e} C},
  journal={University of Wisconsin, Madison, WI},
  pages={1--29},
  year={1998}
}

@article{sangalli_spatial_2013,
    AUTHOR = {Sangalli, Laura M. and Ramsay, James O. and Ramsay, Timothy O.},
     TITLE = {Spatial spline regression models},
     JOURNAL = {Journal of the Royal Statistical Society. Series B. Statistical Methodology},
    VOLUME = {75},
      YEAR = {2013},
    NUMBER = {4},
     PAGES = {681--703},
    doi = {https://doi.org/10.1111/rssb.12009}
}

@article{arnone2019modeling,
  title={Modeling spatially dependent functional data via regression with differential regularization},
  author={Arnone, Eleonora and Azzimonti, Laura and Nobile, Fabio and Sangalli, Laura M},
  journal={Journal of Multivariate Analysis},
  volume={170},
  pages={275--295},
  doi={https://doi.org/10.1016/j.jmva.2018.09.006}, 
  year={2019},
  publisher={Elsevier}
}

@article{des2025exploring,
  title={Exploring nitrogen dioxide spatial concentration via physics-informed multiple quantile regression},
  author={De Sanctis, Marco F and Di Battista, Ilenia and Arnone, Eleonora and Castiglione, Cristian and Palummo, Alessandro and Bernardi, Mauro and Ieva, Francesca and Sangalli, Laura M},
  journal={Environmental and Ecological Statistics,},
  year={2025},
  volume={32}, 
  pages={855--892}, 
  publisher={Springer}, 
  doi={https://doi.org/10.1007/s10651-025-00664-8}
}

@article{ramsay_parameter_2007,
    AUTHOR = {Ramsay, J. O. and Hooker, G. and Campbell, D. and Cao, J.},
     TITLE = {Parameter estimation for differential equations: a generalized smoothing approach},
  JOURNAL = {Journal of the Royal Statistical Society. Series B. Statistical Methodology},
    VOLUME = {69},
      YEAR = {2007},
    NUMBER = {5},
     PAGES = {741--796},
}

@article{xun_parameter_2013,
    AUTHOR = {Xun, Xiaolei and Cao, Jiguo and Mallick, Bani and Maity, Arnab and Carroll, Raymond J.},
     TITLE = {Parameter estimation of partial differential equation models},
  JOURNAL = {Journal of the American Statistical Association},
    VOLUME = {108},
      YEAR = {2013},
    NUMBER = {503},
     PAGES = {1009--1020},
}

@article{craven1978smoothing,
  title={Smoothing noisy data with spline functions: estimating the correct degree of smoothing by the method of generalized cross-validation},
  author={Craven, Peter and Wahba, Grace},
  journal={Numerische mathematik},
  volume={31},
  number={4},
  pages={377--403},
  year={1978},
  publisher={Springer}
}

@article{wahba1985comparison,
  title={A comparison of {GCV} and {GML} for choosing the smoothing parameter in the generalized spline smoothing problem},
  author={Wahba, Grace},
  journal={The annals of statistics},
  pages={1378--1402},
  year={1985},
  publisher={JSTOR}
}

@article{salama_satellite_2022,
  author = {Dina S. Salama and Mohamed Yousif and Yasmine Gedamy and Hanaa M. Ahmed and Mahmoud Ali and Ehab M. Shoukry},
  title = {Satellite observations for monitoring atmospheric {NO}$_2$ in correlation with the existing pollution sources under arid environment},
  journal = {Modeling Earth Systems and Environment},
  volume = {8},
  pages = {4103--4121},
  year = {2022}
}

@article{dibrisco2021,
  title={A spatial mixed-effects regression model for electoral data},
  author={Di Brisco, Agnese Maria and Migliorati, Sonia},
  journal={Statistical Methods \& Applications},
  volume={30},
  number={2},
  pages={543--571},
  year={2021},
  publisher={Springer}
}

@article{lin_zhang1999,
  title={Inference in generalized additive mixed models by using smoothing splines},
  author={Lin, Xihong and Zhang, Daowen},
  journal={Journal of the Royal Statistical Society Series B: Statistical Methodology},
  volume={61},
  number={2},
  pages={381--400},
  year={1999},
  publisher={Oxford University Press}
}

@article{karcher2001generalized,
  title={Generalized nonparametric mixed effects models},
  author={Karcher, Peter and Wang, Yuedong},
  journal={Journal of Computational and Graphical Statistics},
  volume={10},
  number={4},
  pages={641--655},
  year={2001},
  publisher={Taylor \& Francis}
}

@article{smith2003spatiotemporal,
  title={Spatiotemporal modeling of {PM$_{2.5}$} data with missing values},
  author={Smith, Richard L and Kolenikov, Stanislav and Cox, Lawrence H},
  journal={Journal of Geophysical Research: Atmospheres},
  volume={108},
  number={D24},
  year={2003},
  publisher={Wiley Online Library}
}

@article{wood2006low,
  title={Low-rank scale-invariant tensor product smooths for generalized additive mixed models},
  author={Wood, Simon N},
  journal={Biometrics},
  volume={62},
  number={4},
  pages={1025--1036},
  year={2006},
  publisher={Oxford University Press}
}

@article{yanosky2014spatio,
  title={Spatio-temporal modeling of particulate air pollution in the conterminous United States using geographic and meteorological predictors},
  author={Yanosky, Jeff D and Paciorek, Christopher J and Laden, Francine and Hart, Jaime E and Puett, Robin C and Liao, Duanping and Suh, Helen H},
  journal={Environmental Health},
  volume={13},
  number={1},
  pages={63},
  year={2014},
  publisher={Springer}
}

@article{sahu2006spatio,
  title={Spatio-temporal modeling of fine particulate matter},
  author={Sahu, Sujit K and Gelfand, Alan E and Holland, David M},
  journal={Journal of Agricultural, Biological, and Environmental Statistics},
  volume={11},
  number={1},
  pages={61--86},
  year={2006},
  publisher={Springer}
}

@article{damatta2025bayesian,
  title={A Bayesian Spatial-Temporal Functional Model for Data with Block Structure and Repeated Measures},
  author={da Matta, David H and Motta, Mariana R and Garcia, Nancy L and Heinemann, Alexandre B},
  journal={arXiv preprint arXiv:2501.01269},
  year={2025}
}

@article{chen2024long,
  title={Long-term {NO$_2$} exposure and mortality: a comprehensive meta-analysis},
  author={Chen, Xiaoshi and Qi, Ling and Li, Sai and Duan, Xiaoli},
  journal={Environmental Pollution},
  volume={341},
  pages={122971},
  year={2024},
  publisher={Elsevier}
}

@article{wood_soap_2008,
  author = {Simon N. Wood and Mark V. Bravington and Stephan L. Hedley},
  title = {Soap Film Smoothing},
  journal = {Journal of the Royal Statistical Society: Series B (Statistical Methodology)},
  volume = {70},
  pages = {931--955},
  year = {2008}
}

@article{rue2009approximate,
  title={Approximate Bayesian inference for latent Gaussian models by using integrated nested Laplace approximations},
  author={Rue, H{\aa}vard and Martino, Sara and Chopin, Nicolas},
  journal={Journal of the Royal Statistical Society Series B: Statistical Methodology},
  volume={71},
  number={2},
  pages={319--392},
  year={2009},
  publisher={Oxford University Press}
}

@article{trinh2019temperature,
  title={Temperature inversion and air pollution relationship, and its effects on human health in Hanoi City, Vietnam},
  author={Trinh, Thi Thuy and Trinh, Thi Tham and Le, Thi Trinh and Nguyen, The Duc Hanh and Tu, Binh Minh},
  journal={Environmental geochemistry and health},
  volume={41},
  number={2},
  pages={929--937},
  year={2019},
  publisher={Springer}
}

@article{spate,
title = {{spate}: An {R} Package for Spatio-Temporal Modeling with
  a Stochastic Advection-Diffusion Process},
author = {Fabio Sigrist and Hans R. K\"unsch and Werner A. Stahel},
journal = {Journal of Statistical Software},
year = {2015},
volume = {63},
number = {14},
pages = {1--23},
doi = {10.1111/rssb.12061},
}

@article{gamm4,
  title={Package ‘gamm4’},
  author={Wood, Simon and Scheipl, Fabian and Wood, Maintainer Simon},
  journal={Am Stat},
  volume={45},
  number={339},
  pages={0--2},
  year={2017}
}

@article{mgcv,
  title={Package ‘mgcv’},
  author={Wood, Simon and Wood, Maintainer Simon},
  journal={{R} package version},
  volume={1},
  number={29},
  pages={729},
  year={2015}
}

@article{lme4,
    title = {Fitting Linear Mixed-Effects Models Using {lme4}},
    author = {Douglas Bates and Martin M{\"a}chler and Ben Bolker and
      Steve Walker},
    journal = {Journal of Statistical Software},
    year = {2015},
    volume = {67},
    number = {1},
    pages = {1--48}
  }

@article{nlme,
  title={Package ‘nlme’},
  author={Pinheiro, Jos{\'e} and Bates, Douglas and DebRoy, Saikat and Sarkar, Deepayan and Heisterkamp, Siem and Van Willigen, Bert and Maintainer, R},
  journal={Linear and nonlinear mixed effects models, version},
  volume={3},
  number={1},
  pages={274},
  year={2017}
}

@article{anderson2022sdmtmb,
  title={sdmTMB: an {R} package for fast, flexible, and user-friendly generalized linear mixed effects models with spatial and spatiotemporal random fields},
  author={Anderson, Sean C and Ward, Eric J and English, Philina A and Barnett, Lewis AK},
  journal={BioRxiv},
  pages={2022--03},
  year={2022},
  publisher={Cold Spring Harbor Laboratory}
}

@article{Lindgren_2011,
AUTHOR = {Lindgren, Finn and Rue, Håvard and Lindst, Johan},
YEAR = {2011},
VOLUME = {73},
JOURNAL = {Journal of the Royal Statistical Society. Series B. Statistical Methodology},
TITLE = {An explicit link between Gaussian fields and Gaussian Markov random fields: The {SPDE} approach}
}

@article{lindgren2022spde,
  title={The {SPDE} approach for Gaussian and non-Gaussian fields: 10 years and still running},
  author={Lindgren, Finn and Bolin, David and Rue, H{\aa}vard},
  journal={Spatial Statistics},
  volume={50},
  pages={100599},
  year={2022},
  publisher={Elsevier}
}

@article{CLAROTTO2024,
    AUTHOR = {Lucia Clarotto and Denis Allard and Thomas Romary and Nicolas Desassis},
     TITLE = {The SPDE approach for spatio-temporal datasets with advection and diffusion},
   JOURNAL = {Spatial Statistics},
    VOLUME = {62},
     PAGES = {100847},
      YEAR = {2024}
}

@article{wikle-2010-TEST,
    AUTHOR = {Wikle, Christopher K. and Hooten, Mevin},
     TITLE = {A general science-based framework for dynamical spatio-temporal models},
   JOURNAL = {TEST},
    VOLUME = {19},
      YEAR = {2010},
     PAGES = {417--451}
}

@article{richardson-2017,
    AUTHOR = {Richardson, Robert Alan},
     TITLE = {Sparsity in nonlinear dynamic spatiotemporal models using implied advection},
   JOURNAL = {Environmetrics},
    VOLUME = {28},
    NUMBER = {6},
     PAGES = {e2456},
      YEAR = {2017}
}

@article{hefley-2017,
    AUTHOR = {Hefley, Trevor J. and Hooten, Mevin B. and Russell, Robin E. and Walsh, Daniel P. and Powell, James A.},
     TITLE = {When mechanism matters: Bayesian forecasting using models of ecological diffusion},
   JOURNAL = {Ecology Letters},
    VOLUME = {20},
      YEAR = {2017},
    NUMBER = {5},
     PAGES = {640--650}
}

@article{bernardi2017penalized,
  title={A penalized regression model for spatial functional data with application to the analysis of the production of waste in Venice province},
  author={Bernardi, Mara S and Sangalli, Laura M and Mazza, Gabriele and Ramsay, James O},
  journal={Stochastic environmental research and risk assessment},
  volume={31},
  number={1},
  pages={23--38},
  year={2017},
  publisher={Springer},
  doi={https://doi.org/10.1007/s00477-016-1237-3}
}

@article{augustin_spacetime_2013,
  author = {N.H. Augustin and V.M. Trenkel and S.N. Wood and P. Lorance},
  title = {Space‐time modelling of blue ling for fisheries stock management},
  journal = {Environmetrics},
  volume = {24},
  pages = {109--119},
  year = {2013}
}

@article{marra_modelling_2012,
  author = {Giampiero Marra and David L. Miller and Luca Zanin},
  title = {Modelling the spatiotemporal distribution of the incidence of resident foreign population},
  journal = {Statistica Neerlandica},
  volume = {66},
  pages = {133--160},
  year = {2012}
}

@article{o1986automatic,
  title={Automatic smoothing of regression functions in generalized linear models},
  author={O'sullivan, Finbarr and Yandell, Brian S and Raynor Jr, William J},
  journal={Journal of the American Statistical Association},
  volume={81},
  number={393},
  pages={96--103},
  year={1986},
  publisher={Taylor \& Francis}
}

@article{khan2022restricted,
  title={Restricted spatial regression methods: Implications for inference},
  author={Khan, Kori and Calder, Catherine A},
  journal={Journal of the American Statistical Association},
  volume={117},
  number={537},
  pages={482--494},
  year={2022},
  publisher={Taylor \& Francis}
}

@article{carrizo2022general,
  title={A general framework for SPDE-based stationary random fields},
  author={Carrizo Vergara, Ricardo and Allard, Denis and Desassis, Nicolas},
  journal={Bernoulli},
  volume={28},
  number={1},
  pages={1--32},
  year={2022},
  publisher={Bernoulli Society for Mathematical Statistics and Probability}
}

@article{pereira2022geostatistics,
  title={Geostatistics for large datasets on Riemannian manifolds: a matrix-free approach},
  author={Pereira, Mike and Desassis, Nicolas and Allard, Denis},
  journal={arXiv preprint arXiv:2208.12501},
  year={2022}
}

@article{allard2021linking,
  title={Linking physics and spatial statistics: A new family of Boltzmann-Gibbs random fields},
  author={Allard, Denis and Hristopulos, Dionisios T and Opitz, Thomas},
  journal={Electronic Journal of Statistics},
  volume={15},
  number={2},
  pages={4085--4116},
  year={2021},
  publisher={The Institute of Mathematical Statistics and the Bernoulli Society}
}

@book{pinheiro_SandSplus,
  title={Mixed-effects models in S and S-PLUS},
  author={Pinheiro, Jos{\'e} C and Bates, Douglas M},
  year={2000},
  publisher={Springer}
}

@book{wood_Introduction_R,
  title={Generalized additive models: an introduction with R},
  author={Wood, Simon N},
  year={2017},
  publisher={chapman and hall/CRC}
}

@book{Wahba1990SplineMF,
  author = {Grace Wahba},
  title = {Spline Models for Observational Data},
  publisher = {Society for Industrial and Applied Mathematics},
  year = {1990}
}

@incollection{galecki2012linear,
  title={Linear mixed-effects model},
  author={Ga{\l}ecki, Andrzej and Burzykowski, Tomasz},
  booktitle={Linear mixed-effects models using R: a step-by-step approach},
  pages={245--273},
  year={2012},
  publisher={Springer}
}

@misc{fdapde_repo,
    AUTHOR = {Palummo, A. and Arnone, E. and Clemente, A. and Sangalli, L. M. and Ramsay, J. and Formaggia, L.},
    TITLE = {fda{PDE}: Physics-Informed Spatial and Functional Data Analysis},
    HOWPUBLISHED = {GitHub, \url{https://github.com/fdaPDE/fdaPDE}},
    YEAR = {2025}
}

@misc{Geoportale_lombardia_site,
  AUTHOR = {{Regione Lombardia}},
  TITLE = {Geoportale Lombardia},
  YEAR = {2024},
  URL = {https://www.geoportale.regione.lombardia.it/}
}

@misc{WHO2024,
  author       = {{World Health Organization}},
  title        = {Ambient (outdoor) air pollution 2024},
  year         = {2024},
  howpublished = {\url{https://www.who.int/news-room/fact-sheets/detail/ambient-\%28outdoor\%29-air-quality-and-health}}
}

@misc{california_air_sources_board,
  author = {{California Air Resources Board}},
  title = {California Environmental Protection Agency},
  year = {2023},
  url = {https://ww2.arb.ca.gov/}
}

@misc{OpenData_lombardia,
  author = {{Open Data Lombardia, Trasformazione Digitale in Lombardia}},
  title  = {{Regione Lombardia}},
  year   = {2024},
  url    = {https://www.dati.lombardia.it/stories/s/Meteo-inquinamento-aria-e-acqua-e-altri-dati-da-AR/auv9-c2sj}
}

@inproceedings{mullen2008mixed,
  title={Mixed effect and spatial correlation models for analyzing a regional spatial dataset},
  author={Mullen, Randall S and Birkeland, Karl W},
  booktitle={Proceedings of the 2008 International Snow Science Workshop, Whistler, British Columbia},
  pages={421--425},
  year={2008}
}

@misc{INGV,
  author       = {{Istituto Nazionale di Geofisica e Vulcanologia (INGV)}},
  title        = {{National Institute of Geophysics and Volcanology}},
  year         = {2024},
  address      = {Rome, Italy},
  url          = {https://www.ingv.it/}
}

@manual{geoR,
  title        = {geoR: Analysis of Geostatistical Data},
  author       = {Ribeiro Jr, P. J. and Diggle, P. J.},
  year         = {2025},
  note         = {{R} package version 1.9-6},
  doi          = {10.32614/CRAN.package.geoR},
  url          = {https://CRAN.R-project.org/package=geoR}
}

@article{dupont2022spatial,
  title={Spatial+: a novel approach to spatial confounding},
  author={Dupont, Emiko and Wood, Simon N and Augustin, Nicole H},
  journal={Biometrics},
  volume={78},
  number={4},
  pages={1279--1290},
  year={2022},
  publisher={Oxford University Press}
}

@article{ettinger2016spatial,
  title={Spatial regression models over two-dimensional manifolds},
  author={Ettinger, Bree and Perotto, Simona and Sangalli, Laura M},
  journal={Biometrika},
  volume={103},
  number={1},
  pages={71--88},
  year={2016},
  doi={https://doi.org/10.1093/biomet/asv069}, 
  publisher={Oxford University Press}
}

@article{lila2016smooth,
  title={Smooth principal component analysis over two-dimensional manifolds with an application to neuroimaging},
  author={Lila, Eardi and Aston, John AD and Sangalli, Laura M},
  journal={The Annals of Applied Statistics}, 
  doi={10.1214/16-AOAS975}, 
  year={2016}
}

@article{arnone2023analyzing,
  title={Analyzing data in complicated 3D domains: Smoothing, semiparametric regression, and functional principal component analysis},
  author={Arnone, Eleonora and Negri, Luca and Panzica, Ferruccio and Sangalli, Laura M},
  journal={Biometrics},
  volume={79},
  number={4},
  pages={3510--3521},
  year={2023},
  doi={https://doi.org/10.1111/biom.13845}, 
  publisher={Wiley Online Library}
}

@article{clemente2026nonparametric,
  title={Nonparametric estimators over metric graphs},
  author={Clemente, Aldo and Arnone, Eleonora and Mateu, Jorge and Sangalli, Laura M},
  journal={Biometrika},
  pages={asag029},
  year={2026},
  doi={https://doi.org/10.1093/biomet/asag029}, 
  publisher={Oxford University Press}
}

\AddToHook{enddocument/afteraux}{%
\immediate\write18{
cp output.aux elsarticle-template-num.aux
}%
}

\end{document}